\documentclass[journal=jctcce,manuscript=article]{achemso}

\usepackage{mathtools,amssymb,graphicx}
\usepackage[plainpages=false,pdfpagelabels,colorlinks=true,linkcolor=red,urlcolor=blue,citecolor=blue,pdftitle={Title},pdfauthor={},pdfdisplaydoctitle=true,pdfduplex=DuplexFlipL†ongEdge]{hyperref}
\usepackage{soul} 
\usepackage[dvipsnames,usenames]{color}
\usepackage[normalem]{ulem}
\usepackage{tabularx}
\usepackage{graphicx}
\usepackage{xcolor}
\usepackage{bbold, verbatim} 
\usepackage{float}
\graphicspath{ {./figures/} }
\usepackage{tabularx}
\usepackage{rotating}
\usepackage{braket}
\usepackage{multirow}
\usepackage{ulem}
\usepackage{amsmath}

\newcommand{\turborvb}{\textsc{TurboRVB}} 
\newcommand{\qmcpack}{\textsc{QMCPACK}}
\newcommand{\pyscf}{\textsc{PySCF}} 
 
\newcommand{\trexio}{\textsc{TREX-IO}}

\makeatletter
\def\Hline{
\noalign{\ifnum0=`}\fi\hrule \@height 1pt \futurelet
\reserved@a\@xhline}
\makeatother

\author{Kousuke Nakano}
\email{kousuke_1123@icloud.com}
\affiliation[NIMS]
{Center for Basic Research on Materials, National Institute for Materials Science (NIMS), Tsukuba, Ibaraki 305-0047, Japan}

\author{Benjamin X. Shi}
\affiliation[CAM]
{Yusuf Hamied Department of Chemistry, University of Cambridge, Cambridge CB2 1EW, United Kingdom}

\author{Dario Alf{\`e}}
\affiliation[UNiNa]{Dipartimento di Fisica Ettore Pancini, Universit\`a di Napoli Federico II, Monte S. Angelo, I-80126 Napoli, Italy}
\alsoaffiliation[UCL]{Department of Earth Sciences, University College London, Gower Street, London WC1E 6BT, United Kingdom}
\alsoaffiliation[TYC]{Thomas Young Centre and London Centre for Nanotechnology, 17-19 Gordon Street, London WC1H 0AH, United Kingdom}

\author{Andrea Zen}
\email{andrea.zen@unina.it}
\affiliation[UNiNa]{Dipartimento di Fisica Ettore Pancini, Universit\`a di Napoli Federico II, Monte S. Angelo, I-80126 Napoli, Italy}
\alsoaffiliation[UCL]{Department of Earth Sciences, University College London, Gower Street, London WC1E 6BT, United Kingdom}

\title{Basis set incompleteness errors in fixed-node diffusion Monte Carlo calculations on non-covalent interactions} 

\begin{document}

\begin{tocentry}
\centering
\includegraphics[width=8.0cm]{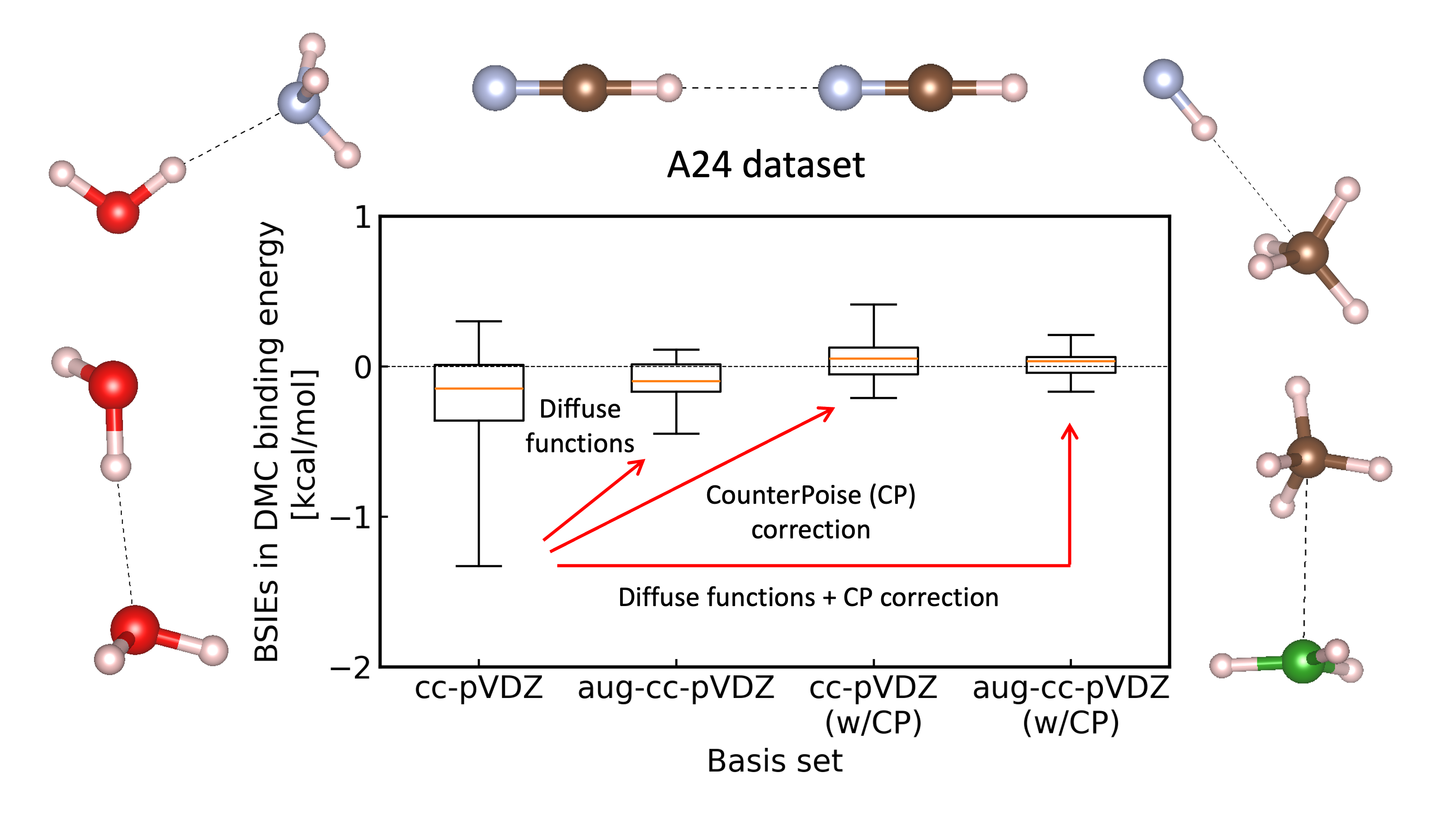}
\label{For Table of Contents Only}

\end{tocentry}

\begin{abstract}
Basis set incompleteness error (BSIE) is a common source of error in quantum chemistry (QC) calculations, but it has not been comprehensively studied in fixed-node Diffusion Monte Carlo (FN-DMC) calculations. 
FN-DMC, being a projection method, is often considered minimally affected by basis set biases. 
Here, we show that this assumption is not always valid.
While the relative error introduced by a small basis set in the total FN-DMC energy is minor, it can become significant in binding energy ($E_{\rm b}$) evaluations of weakly interacting systems. We systematically investigated BSIEs in FN-DMC-based binding energy ($E_{\rm b}$) evaluations using the A24 dataset, a well-known benchmark set of 24 non-covalently bound dimers. Contrary to common expectations, we found that BSIEs in FN-DMC evaluations of $E_{\rm b}$ are indeed significant when small localized basis sets, such as cc-pVDZ, are employed.
We observed that BSIEs are larger in dimers with hydrogen-bonding interactions and smaller in dispersion-dominated interactions. 
We also found that augmenting the basis sets with diffuse orbitals, using counterpoise (CP) correction, or both, effectively mitigates BSIEs.
\end{abstract}



\section{Introduction}
\label{sec:intro}

Diffusion Monte Carlo (DMC)~{\cite{1986CEP_GFMC, 2001FOU_qmcreview}} is a state-of-the-art electronic structure method used for predicting and understanding phenomena in materials science, chemistry, and physics.
In particular, DMC can achieve highly accurate quantitative predictions, typically surpassing those of mean-field approaches like density functional theory (DFT).
This level of accuracy has proven essential for studying systems challenging for DFT, such as high-pressure hydrogen~{\cite{2015DRU_hydrogen, 2018MAZ_hydrogen, 2022TIR_hydrogen_MLP, 2022_LY_hydrogen, 2023NIU_hyrdogen_MLP, 2023MON_hydrogen, 2024GIA_hydrogen_MLP}}, layered materials~{\cite{2020KRO_TiS2, 2021ICH_CrI3, 2022NIK_BNs, 2023WIN_CrX3, 2023WIN_VSe2}}, molecular crystals~{\cite{2018ZEN_MolCrys, 2024FLA_DMC_MolCrys}}, and molecular adsorption on surfaces~{\cite{2008BEA_absorption, 2016ZEN_absorption, 2017HAM_absorption, 2019CHE_absorption, 2023BEN_absorption}.

In theory, DMC is an exact technique to project the ground state (GS) of a Hamiltonian. 
However, in practical applications to fermionic systems (e.g., atoms, molecules, and materials), it relies on the fixed-node (FN) approximation to maintain the antisymmetry of the wavefunction. 
The FN approximation constrains the nodal surface of the projected state to that of a trial wavefunction, which can be generated by methods such as DFT, Hartree-Fock (HF), or correlated quantum chemistry (QC) methods, including the complete active space self consistent field (CASSCF) method.

The approaches used to generate the trial wavefunction are not exact, so its nodal surface is not exact either, yielding an error on the FN-DMC evaluations called the FN error. 
The closer the nodal surface of the trial wavefunction to the nodal surface of the exact GS, the smaller the FN error.
There are other approximations in FN-DMC, but typically the major source of error is the FN error.
The FN error depends on the accuracy of the trial wavefunction, which in turn depends on the level of theory employed to generate it (e.g., we expect a CASSCF wavefunction to have a better nodal surface than a DFT or an HF wavefunction) and on the completeness of the employed \textit{basis set} representation (the larger the basis set, the better is the wavefunction and usually its nodal surface).
While previous work has explored the influence of trial wavefunction accuracy,\cite{2011KOR, 2010NEM_G2set, 2012JCP_Petruzielo_DMC-G2, 2013JCTC_Zen_water, Scemama2016, 2016MIC, 2020JCP_Scemama_FNDMC-rangesep}
the impact of the basis set incompleteness errors (BSIEs) in FN-DMC has yet to be comprehensively and systematically explored in the context of non-covalent interaction evaluations, which is one of the most prominent applications of FN-DMC~\cite{2008KOR_S22set, 2010NEM_G2set, 2013DUB, 2014DUB, 2015PRL_bilayergraphene, 2019NAK_vdw, Zen2018_PNAS, 2020BEN_L7set, 2021YAS_C60_CPPA, 2020KRO_TiS2, 2021ICH_CrI3, 2022NIK_BNs, 2023WIN_CrX3, 2023WIN_VSe2, 2023RAG_G2set, 2023BEN_absorption}.

In QC and DFT methods, BSIEs are a dominant error source that requires careful control, 
yet it has been often assumed that FN-DMC is relatively immune to BSIEs from the trial wavefunction~{\cite{Dubecky2016}} because it depends only on the nodal surface, not on the full wavefunction amplitude.
In this work, we systematically investigate how these assumptions hold up by analyzing BSIEs in FN-DMC calculations.

BSIEs are especially pronounced in QC and DFT methods when describing non-covalent interactions. In this context, the quantity of interest is typically the binding energy of a dimer complex (AB), defined as:
\begin{equation}
\label{eq:def-Eb}
{E_{\rm b}} = {E^{\textrm{AB}}} - E^{\rm{A}} - E^{\rm{B}},
\end{equation}
where $E^{\rm{A}}$, $E^{\rm{B}}$, and ${E^{\rm{AB}}}$ are the total energies of monomer A, monomer B, and the AB dimer complex, respectively.
This study focuses on the propagation of BSIEs from the trial wavefunction in FN-DMC calculations of $E_{\rm b}$, a particularly relevant area of investigation given the high sensitivity of non-covalent interactions to basis set quality~\cite{2021YAS_C60_CPPA, 2024NAK_JAGPn, 2024SCH_CCSDTQ}.

A basis set consists of a number of basis functions that are used to represent the electronic wave function, with the complete basis set (CBS) limit achieved when expanded towards an (infinite) set of functions.
The BSIE is the deviation from the CBS limit~{\cite{1994DUI, 2000DUN} and for a binding energy $E_\textrm{b}$, it is defined as~{\cite{1994DUI, 2000DUN}}:
\begin{equation}
\label{eq:def-bsie}
E_{\rm b}^{\rm BSIE}(M^\textrm{A}, M^\textrm{B}, M^\textrm{AB}) = {{E_{\rm b}}(M^\textrm{A}, M^\textrm{B}, M^\textrm{AB})} - E_{\rm b}^{\rm CBS},
\end{equation}
where $M^\textrm{A}$, $M^\textrm{B}$ and $M^\textrm{AB}$ denote the number and type of basis functions employed in the calculation of $E^\textrm{A}$, $E^\textrm{B}$ and $E^\textrm{AB}$ respectively within Eq.~\ref{eq:def-Eb}, and 
$E_{\rm b}^{\rm CBS}$
denotes the binding energy in the CBS limit.
Two common choices of basis function types are plane waves (PWs) and atom-centered Gaussian Type Orbitals (GTOs).
On the one hand, BSIEs are well-controlled with PWs because systematic convergence towards the CBS limit can be achieved by monotonically increasing the kinetic (i.e., cut-off) energy of the included PWs.
On the other hand, errors in GTOs are less well-behaved, with users selecting from `families' of available basis sets consisting of increasing sizes, often denoted by the number of `zeta' basis functions per occupied valence orbital.
A popular example is the correlation consistent basis-set family, developed by Dunning and coworkers~{\cite{1989DUN}}, for instance the correlation-consistent polarized valence $n$-zeta (cc-pV$n$Z), where $n$, the cardinal number, can take on double (D), triple (T), quadruple (Q), quintuple (5) and sextuple (6) zeta functions on each atom.
It is also common to augment these with additional diffuse functions, which are denoted by an `aug-' prefix in front.

When using GTOs, or any other set of atom-centered basis functions, to compute binding energies, it is crucial to distinguish BSIEs from basis-set superposition errors (BSSEs)~{\cite{1994DUI, 2000DUN}}, a related source of error.
BSSE occurs when basis functions of interacting molecular systems A and B in the AB dimer overlap,
increasing the variational space for the AB dimer with respect to the A and B monomers, thus leading to an overestimation of $E_\textrm{b}$.\footnote{Note that PW basis sets are not affected by any BSSE, while they can be affected by a BSIE when the PW cutoff is too small. }
This error is defined by Boys and Bernardi~{\cite{1970BOY}} as:
\begin{equation}
\label{eq:def-bsse}
{E_{\rm b}^{\rm BSSE}(M^\textrm{A}, M^\textrm{B}, M^\textrm{AB})} = [E^{\rm{A}}(M^\textrm{AB}) - E^{\rm{A}}(M^\textrm{A})] + [E^{\rm{B}}(M^\textrm{AB}) - E^{\rm{B}}(M^\textrm{B})],
\end{equation}
involving two separate calculations on each monomer.
For monomer A, alongside the original basis set $E^{\rm{A}}(M^\textrm{A})$, a calculation including additional empty `ghost' functions from monomer B is also performed to get $E^{\rm{A}}(M^\textrm{AB})$, as proposed by Boys and Bernardi~{\cite{1970BOY}}.
The difference between the two quantities, appearing in Eq.~\ref{eq:def-bsse}, then provides an estimate on the effect of the basis set superposition on the energy of each monomer. 
Thus, the BSSE error ${E_{\rm b}^{\rm BSSE}}$ can be used to correct the original $E_\textrm{b}$ evaluation to obtain a counterpoise (CP) corrected estimate of the binding energy: $E_{\rm b}^{\rm CP} = E_{\rm b} - E_{\rm b}^{\rm BSSE}$.
It must be emphasized that the CP corrected estimates still suffer from BSIE, although they are typically closer to the CBS limit~{\cite{2000DUN}}, and typically underbind $E_\textrm{b}$.\footnote{Note that $E_{\rm b}^{\rm CP} \ge E_{\rm b}$, because $E_{\rm b}^{\rm BSSE} \le 0$ as $M^{\rm AB}>M^{\rm A}$ and $M^{\rm AB}>M^{\rm B}$.}
In the CBS limit, both BSIE and BSSE will vanish.

To date, only a few studies have reported BSIEs in FN-DMC for $E_\textrm{b}$ calculations of non-covalent interactions and to our knowledge, none have studied the effect of CP corrections.
Korth et al.~{\cite{2011KOR}} reported the difference between non-covalent interaction energies of the Li-thiophene complex obtained with cc-pVTZ and cc-pVQZ basis sets. The results from the cc-pVQZ basis were close to the CCSD(T)/CBS reference value. 
Dubeck\'{y} et al.~{\cite{2013DUB}} studied the eﬀect of the cardinal number $n$ and augmentation functions in ammonia dimer. On the one hand, they revealed that the higher cardinality number $n$ (from cc-pVTZ to cc-pVQZ) has a smaller eﬀect on the overall accuracy than the augmentation does. On the other hand, the additional diffuse functions (aug-) were found to be crucial to reach the reference CCSD(T)/CBS interaction energy value because the augmentation functions likely improve the tails of trial wavefunctions that are crucial for describing van der Waals complexes correctly. They recommended the aug-cc-pVTZ basis set as the most reasonable choice with respect to the price/performance ratio.
Very recently, \citet{2024JCP_Zhou_WatClusters} evaluated barrier heights and complexation energies in small water, ammonia, and hydrogen fluoride clusters using FN-DMC with basis sets of increasing completeness, and recommend basis sets containing diffuse basis functions.

In this paper, we present a detailed analysis of the basis set effects, BSIEs and BSSEs, in DMC binding energy calculations, specifically focusing on non-covalent interactions. Our findings indicate that while BSIEs and BSSEs in FN-DMC are substantially reduced compared to those in the trial wave function, they are not negligible. The key conclusions to get CBS-limit binding energies (i.e., negligible BSIEs and BSSEs) from our work are: (1) cc-pVDZ is sufficient when CP correction is applied and (2) the aug-cc-pVTZ basis set performs well without the need for CP correction.

\section{Computational details}
\label{sec:methos}
To investigate BSIEs in DMC calculations systematically, we computed binding energies ($E_{\rm b}$) of the complex systems included in the A24 dataset~{\cite{2013REZ}}. The A24 dataset is a set of non-covalently bound dimers, consisting of systems dominated by H-bonding, dispersion and a mixture of both~{\cite{2013REZ}}. The dataset was intended to test the accuracy of computational methods that are used as benchmarks in larger model systems. 
We employed the correlation consistent (cc) GTOs accompanied by the correlation consistent effective core potentials~{\cite{2017BEN,2018BEN}} (ccECP) in this study. 
The majority of the QMC results reported in this work are obtained using the \turborvb~{\cite{2020NAK_turborvb}} ab-initio QMC packages. 
\turborvb\ performs QMC calculations using trial wave functions expressed in terms of localized atomic orbitals, such as GTOs.
\turborvb\ also implements the CP correction for QMC calculations using trial wavefunctions with GTOs, allowing one to study both BSIEs and BSSEs. 

\turborvb\ implements the lattice discretized version of the FN-DMC calculations (LRDMC)~{\cite{2005CAS_lrdmc, 2020NAK_turborvb}}. 
Notice that the infinitesimal mesh limit of LRDMC evaluations is equivalent to the infinitesimal timestep limit in standard DMC evaluations, provided that the computational setup (i.e., trial wave function, pseudopotential, localization approximation of the non-local pseudopotential terms) is the same.
The LRDMC calculations with \turborvb\ were performed by the single-grid scheme~{\cite{2005CAS_lrdmc}} with lattice spaces $a$ = 0.30, 0.25, 0.20, and 0.10 Bohr. 
BSSEs were computed at each lattice space according to eq.~{\ref{eq:def-bsse}}, and then the obtained values were extrapolated to $a \rightarrow 0$ using $E_{\rm b}^{\rm BSSE}(a^2)=k_2 \cdot a^2 + E_{\rm b}^{\rm BSSE}$, where $E_{\rm b}^{\rm BSSE}$ is the extrapolated BSSE. 
In computing BSIEs, the binding energies computed with the aug-cc-pV6Z were used as the reference values, i.e., $E_{\rm b}^{\rm CBS}$ in eq.~{\ref{eq:def-bsie}}, for each complex system because, as shown in the following section, the aug-cc-pV6Z basis has reached the CBS limit. 
The binding energy obtained with each basis set was extrapolated to $a \rightarrow 0$ using $E_{\rm b}(a^2)=k_4 \cdot a^4 + k_2 \cdot a^2 + E_{\rm b}$, where $E_{\rm b}$ is the extrapolated binding energy, and then BSIEs were computed according to eq.~{\ref{eq:def-bsie}}.

The ccECP pseudopotentials are semi-local effective core potentials, as with most available pseudopotentials, so the DMC results depend on how the sign problem from its non-local term is addressed. 
In this study, we used the determinant locality T-move (DTM)\cite{2019ZEN} scheme in the majority of the calculations shown here, which are performed with \turborvb. 

%
For the DFT calculations that generate trial wavefunctions with GTOs for subsequent QMC calculations via \trexio~{\cite{2023POS}} files, we used the \pyscf~{\cite{2018SUN, 2020SUN}} package, with the PZ-LDA~{\cite{1981PER}} exchange-correlation functional. 
For LRDMC calculations with \turborvb\, the obtained trial wavefunctions are combined with the two-body and the three-body Jastrow factors~{\cite{2020NAK_turborvb}}. The three-body Jastrow factors are not attached to the ghost atoms in CP calculations. The parameters in the Jastrow factors were optimized using the Stochastic Reconfiguration method~{\cite{1998SOR}}. We notice that the optimization of the Jastrow factor does not affect the extrapolated LRDMC total and binding energies since the DTM is employed in this study. In this sense, the obtained conclusions in this study are deterministic.

In the SI, we also compare QMC evaluations obtained using PW basis sets in comparison with localized GTO basis sets. 
The comparison uses results obtained with the \qmcpack\ package~{\cite{2018KIM_qmcpack, 2020KEN_qmcpack}}, which implements wavefunctions using either PW or GTO basis sets.
Details about the \qmcpack\ calculations are provided in section S1 of the SI.


%

\section{Basis-set convergence checks to estimate the binding energies in the CBS limit}
\label{sec:results-BSSEs}

To estimate BSIEs, the binding energies in the CBS limit are needed, as described in eq.~{\ref{eq:def-bsie}}. Since zero BSSE implies zero BSIE in binding energy calculation, computing BSSEs is helpful to decide which basis set should be used to compute $E_{\rm b}^{\rm CBS}$ in eq.~{\ref{eq:def-bsie}}. 

Figure.~{\ref{fig:BSSE-summary-turbo}} (a) shows BSSEs in the binding energies of the A24 set computed by LRDMC implemented in \turborvb. They were obtained using cc-pVDZ, cc-pVTZ, aug-cc-pVTZ, and aug-cc-pV6Z basis sets. 
Figure.~{\ref{fig:BSSE-summary-turbo}} (b) shows the violin plots of the BSSEs. The figures reveal that the binding energies obtained with the cc-pVDZ and cc-pVTZ basis sets have significant BSSEs, indicating that the small basis sets are far from the CBS limit. 
BSSEs vanish for all molecules with the aug-cc-pV6Z basis set within an interval of three standard deviations ($\pm 3 \sigma$, corresponding to a confidence of 99.7\%), indicating that the aug-cc-pV6Z basis set has reached the CBS limit.

\begin{figure*}[htbp]
\centering
\includegraphics[width=1.0\textwidth]{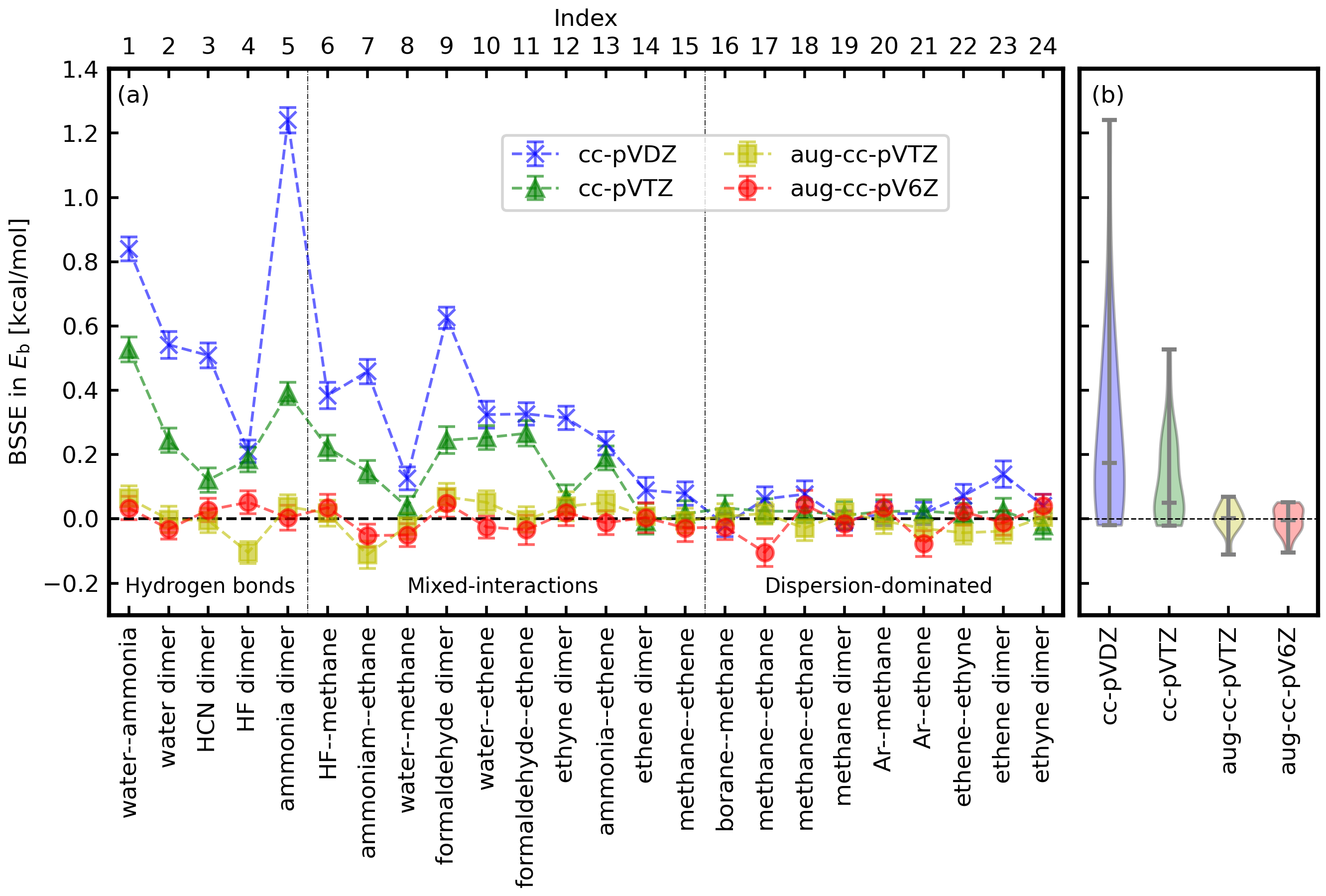}
\caption{(a) The BSSEs in the binding energies of the A24 set computed by LRDMC with cc-pVDZ, cc-pVTZ, aug-cc-pVTZ, and aug-cc-pV6Z basis sets. The plotted BSSEs are the values extrapolated to the infinitesimal lattice space. The error bars represent $1\sigma$. (b) The violin plots for the obtained BSSEs, with median of the distribution indicated with a grey line inside the violin plot.
}
\label{fig:BSSE-summary-turbo}
\end{figure*}

In addition, to double-check that the aug-cc-pV6Z basis set gives the CBS-limit binding energies, we computed the DMC binding energies on all A24 dimers using aug-cc-pV6Z, as well as smaller GTO basis sets, and PW basis sets with a very large cutoff. We made this calculations using \qmcpack, which allows to use both localized and PW basis sets. The binding energies values are reported in Table~{\color{blue}S1} of the SI, and a comparison between the evaluations with different basis sets is shown in Figure.~{\color{blue}S1} of the SI.
The results indicate that aug-cc-pV6Z and large-cutoff-PW basis sets give consistent binding energies within an interval of $\pm 3 \sigma$, supporting the above argument that the aug-cc-pV6Z basis set gives converged binding energies.

Therefore, both the BSSEs evaluation and the comparison with large-cutoff-PW indicate that the aug-cc-pV6Z has reached the CBS limit.
Thus, we can use the binding energies obtained with the aug-cc-pV6Z basis sets (without the CP correction) as reference values (i.e., $E_{\rm b}^{\rm CBS}$ in eq.~{\ref{eq:def-bsie}}) in the following BSIE analysis.
The reference DMC values are reported in Table~\ref{tab:ccsdt-dmc-comparison}.

\begin{center}
\begin{table}[hbtp]
\caption{\label{tab:ccsdt-dmc-comparison}
The binding energies $E_{\rm b}$, in kcal/mol, of the 24 molecular dimers contained in the A24 dataset~\cite{2013REZ}. 
$E_{\rm b}^{\rm DMC}$ column shows results obtained in this work, from LRDMC calculations employing the ccECP pseudopotentials~\cite{2017BEN,2018BEN} with the DTM approximation~\cite{2019ZEN}, and a trial wavefunction with the determinant from a LDA-PZ DFT calculation, constructed with ccecp-aug-cc-pV6Z basis sets. $E_{\rm b}^{\rm CCSD(T)}$ column shows the evaluations from \citet{2013REZ}, computed by CCSD(T) with extrapolations to the CBS limits. The last column shows the differences $\Delta = E_{\rm b}^{\rm CCSD(T)} - E_{\rm b}^{\rm DMC}$ between the LRDMC and CCSD(T) values, with the root mean square deviation RMSD at the end.
}
\begin{tabular}{c|ccc}
\Hline
Label & $E_{\rm b}^{\rm DMC}$ & $E_{\rm b}^{\rm CCSD(T)}$ & $\Delta$ \\
\Hline
water{-}{-}ammonia & -6.75(7) & -6.493 & 0.26(7) \\
water dimer & -5.10(8) & -5.006 & 0.09(8) \\
HCN dimer & -5.09(7) & -4.745 & 0.34(7) \\
HF dimer & -4.74(7) & -4.581 & 0.16(7) \\
ammonia dimer & -3.10(6) & -3.137 & -0.04(6) \\
HF{-}{-}methane & -1.64(7) & -1.654 & -0.01(7) \\
ammonia{-}{-}methane & -0.80(7) & -0.765 & 0.04(7) \\
water{-}{-}methane & -0.58(6) & -0.663 & -0.08(6) \\
formaldehyde dimer & -4.42(9) & -4.554 & -0.13(9) \\
water{-}{-}ethene & -2.50(10) & -2.557 & -0.06(10) \\
formaldehyde{-}{-}ethene & -1.71(10) & -1.621 & 0.09(10) \\
ethyne dimer & -1.44(7) & -1.524 & -0.08(7) \\
ammonia{-}{-}ethene & -1.38(6) & -1.374 & 0.01(6) \\
ethene dimer & -0.97(9) & -1.090 & -0.12(9) \\
methane{-}{-}ethene & -0.56(6) & -0.502 & 0.06(6) \\
borane{-}{-}methane & -1.46(7) & -1.485 & -0.03(7) \\
methane{-}{-}ethane & -0.65(9) & -0.827 & -0.18(9) \\
methane{-}{-}ethane & -0.57(8) & -0.607 & -0.04(8) \\
methane dimer & -0.58(6) & -0.533 & 0.05(6) \\
Ar{-}{-}methane & -0.36(8) & -0.405 & -0.05(8) \\
Ar{-}{-}ethene & -0.24(7) & -0.364 & -0.12(7) \\
ethene{-}{-}ethyne & 1.04(9) & 0.821 & -0.22(9) \\
ethene dimer & 1.04(8) & 0.934 & -0.11(8) \\
ethyne dimer & 1.32(8) & 1.115 & -0.21(8) \\
\Hline
RMSD  & --- & --- & 0.135 \\
\Hline
\end{tabular}
\end{table}
\end{center}

\section{Bias against the binding energies in the CBS limit}
\label{sec:results-BSIEs}

\begin{figure*}[htbp]
\centering
\includegraphics[width=1.0\textwidth]{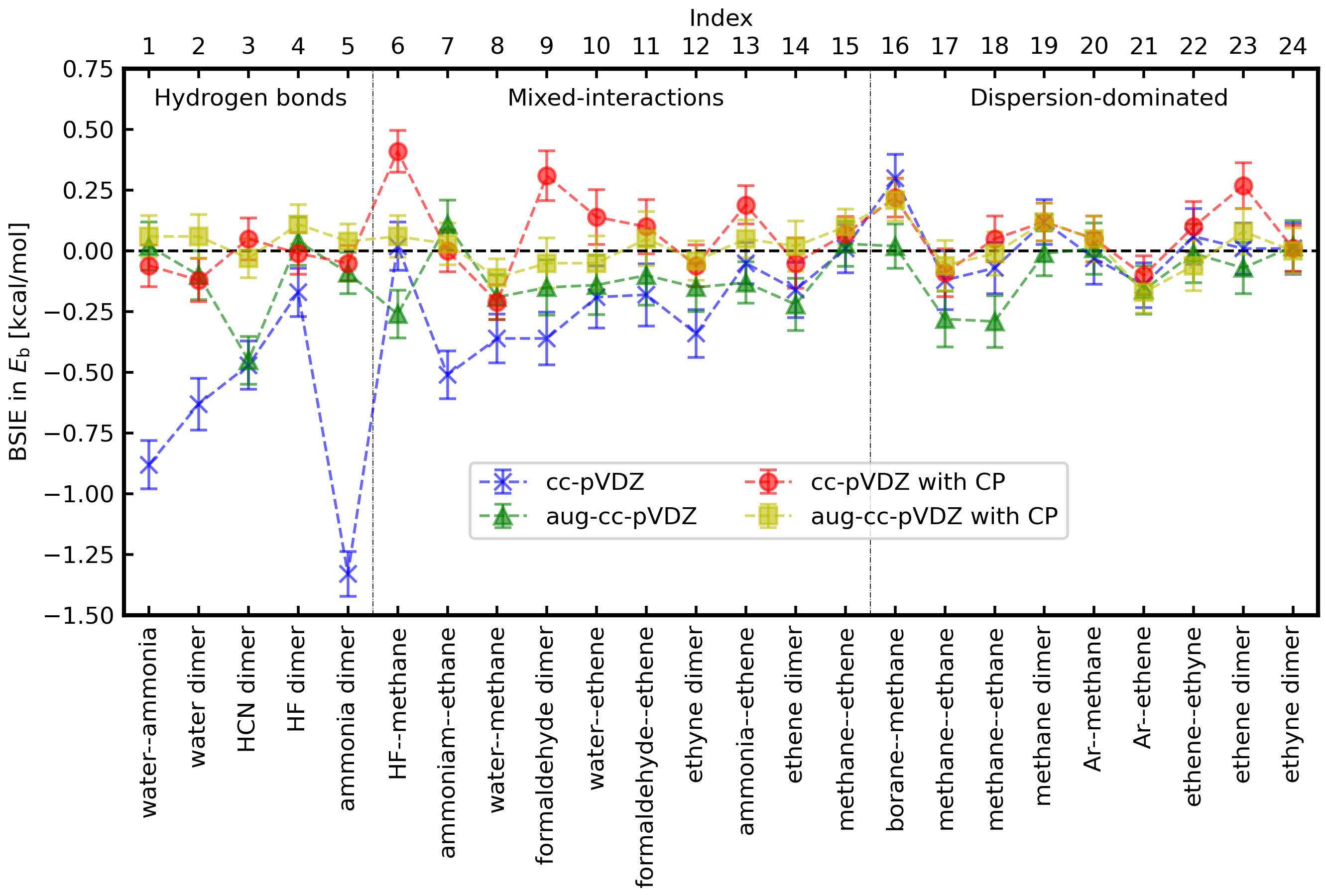}
\caption{BSIEs in the binding energies of the A24 set, estimated from LRDMC calculations in the limit of infinitesimal lattice spaces. The error bars represent $1\sigma$.}
\label{fig:BSIE-summary-turbo}
\end{figure*}


BSIEs in binding energies obtained from LRDMC calculations with the cc-pVDZ and aug-cc-pVDZ basis sets (for the ccECPs~{\cite{2017BEN,2018BEN}}) with and without the CP corrections are shown in Figure~\ref{fig:BSIE-summary-turbo} for each of the 24 dimers of the A24 dataset.
The distribution of the BSIEs across the dataset for the same basis set and CP correction combinations is shown in Figure~\ref{fig:BSIE-summary}(a) via a violin plot.
By comparison, BSIEs in MP2 calculations are shown in Figure~\ref{fig:BSIE-summary}(b).

In Fig.~{\ref{fig:BSIE-summary-turbo}}, the comparison between the BSIEs with cc-pVDZ (without CP) and with aug-cc-pVDZ (without CP) reveals that the augmentation of the basis set drastically decreases BSIEs, specifically for the complex systems with hydrogen-bond interactions. The most significant discrepancy is seen for the ammonia dimer, for which Dubeck\'{y} et al.~{\cite{2013DUB}} also reported that the additional diffuse functions (i.e., augmentation) were crucial to reach the reference CBS interaction energy value. They interpreted the outcome such that augmentation functions likely improve the tails of trial wavefunctions that are crucial for describing the weak interactions correctly~{\cite{2013DUB}}.
%
The wider set of results reported in this work supports the above interpretation.
The interaction among molecules included in the A24 dataset are categorized into three groups~{\cite{2013REZ}}: Hydrogen bonds (index 1 to 5), mixed interactions (index 6 to 15), and dispersion-dominated interactions (index 16 to 24). 
Dimers in the hydrogen-bond group show the most significant BSIEs, while the dispersion-dominated dimers are less affected by BSIEs. 
%
%
The hydrogen bond, which originates from the Coulomb interactions, has the long-tail effect (e.g., $1/r^2$) compared with the dispersion-dominated ones, which are typically shorter-range interactions (e.g., $1/r^6$). 
It appears that the long-tail of the interaction has an effect on the nodal surface (affecting the FN-DMC evaluations), which can be improved if {\it diffuse} functions are available in the basis set.

%
In Fig.~{\ref{fig:BSIE-summary-turbo}}, the comparison between BSIEs with cc-pVDZ with and without the CP correction of the basis set shows that this correction alleviates the BSIEs. It implies that the basis sets assigned to the ghost atoms can compensate missing diffuse functions in the cc-pVDZ basis set, thus improving the nodal surface of the monomers and decreasing the FN error on the binding energy evaluations. 
This suggests that the CP correction is an alternative way to eliminate BSIEs in DMC calculations.
The simultaneous use of augmentation and CP leads to a synergistic effect, as can be appreciated in Fig.~\ref{fig:BSIE-summary-turbo} observing the evaluations obtained using aug-cc-pVDZ with CP.

\begin{figure*}[htbp]
\centering
\includegraphics[width=1.0\textwidth]{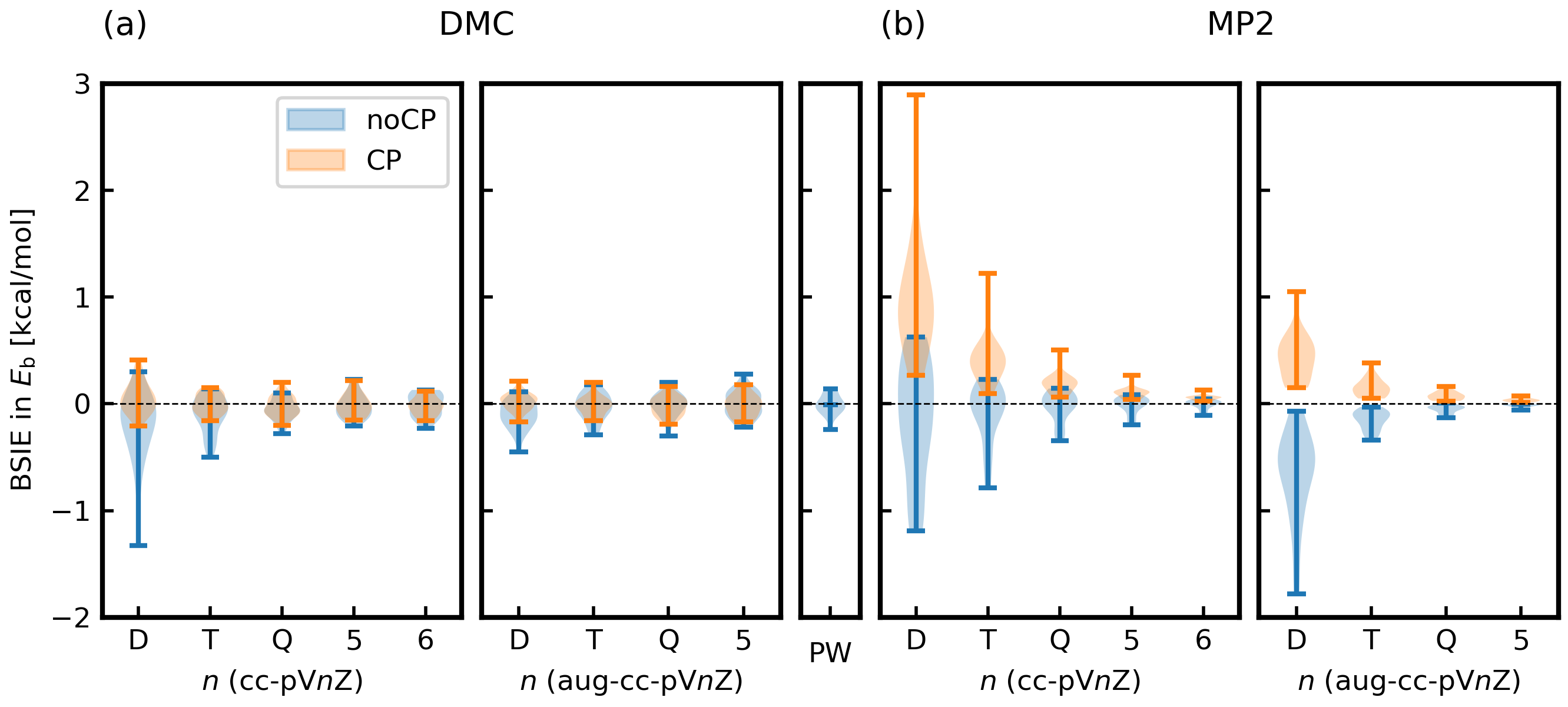}
\caption{The violin plots of BSIEs in the binding energy calculations of the A24 data set with and without the CP corrections. (a) LRDMC with cc-pV$n$Z ($n$=D,T,Q,5,6) and aug-cc-pV$n$Z ($n$=D,T,Q,5) and DMC with PW. (b) MP2 with cc-pV$n$Z ($n$=D,T,Q,5,6) and with aug-cc-pV$n$Z ($n$=D,T,Q,5). The reference binding energies are those obtained with aug-cc-pV6Z basis without CP correction. 
}
\label{fig:BSIE-summary}
\end{figure*}

%
Figure~{\ref{fig:BSIE-summary}} (a) summarizes the BSIEs obtained with all the family members of the cc basis sets and PW used in this study. The left panel of Fig.~{\ref{fig:BSIE-summary}} (a) plots the BSIEs with the non-augmented cc basis sets (cc-pV$n$Z: $n$=D,T,Q,5,6), revealing that, to get binding energies in the CBS limit within their statistical errors ($3 \sigma \sim 0.25$ kcal/mol), one needs the cc-pVQZ without the CP corrections or the cc-pVTZ with the CP correction. The central panel of Fig.~{\ref{fig:BSIE-summary}} (a) plots BSIEs with the augmented cc basis sets (aug-cc-pV$n$Z: $n$=D,T,Q,5), indicating that the augmentations of the basis sets improve the situation. To get binding energies in the CBS limit within their statistical errors, one needs the aug-cc-pVTZ without the CP correction or the aug-cc-pVDZ basis with the CP correction. The right panel of Fig.~{\ref{fig:BSIE-summary}} (a) plots BSIEs with PW basis set, confirming that aug-cc-pV6Z basis set gives binding energies in the CBS limit (i.e., zero BSIEs within the statistical errors).

It is informative to compare the BSIEs obtained by DMC with those obtained using a quantum chemistry method, such as MP2, to understand the impact of basis sets. The comparison between Figs.~{\ref{fig:BSIE-summary}} (a) and (b) reveals that BSIEs in the DMC calculations are not as significant as in the MP2 calculations, as believed in the QMC community. This is true not only for the binding energies, but also the total energies of fragments and complexes, as shown in Figs.~{\color{blue}S2}. In panel (b), the asymptotic behaviors with $n$ are seen in the binding energies computed by MP2. For QC calculations, the most common and established procedure to reach the CBS limit is the extrapolation of the binding energies with consecutive cardinal numbers~\cite{neeseRevisitingAtomicNatural2011}. The asymptotic behaviors (c.f., the binding energies of all dimers are shown in SI) allow the extrapolation, and, in fact, the CP correction in MP2 calculations is necessary for smoother extrapolations to the CBS limit, as mentioned in Ref.~{\citenum{2000DUN}}. Instead, in DMC calculations, the extrapolation is no longer needed when the CP correction is applied. One can get the binding energy in the CBS limit only with a {\it single} medium-size basis set (such as TZ and QZ), which helps decrease the computational cost to reach the CBS limit.

%
In summary, we revealed that both BSSE and BSIE are not negligible in DMC binding energy calculations if one targets to compute binding energies of complex systems within the sub-chemical accuracy (i.e., $\sim$ 0.1 kcal/mol). The augmentation (i.e., more diffuse functions) of a basis set and the CP correction for a basis set are both helpful to eliminate BSIEs, i.e., to get binding energies in the CBS limit.

\section{A24 benchmark test revisited}
\label{sec:results-A24}
Benchmarks for the A24 set were done by Dubeck\'{y} et al.~{\cite{2014DUB}} and by Nakano et al.~{\cite{2023NAK}} with the aug-TZV basis sets associated with the ECPs developed by Burkatzki et al.~{\cite{2007burkatzkiecp}} and the cc-pVTZ basis sets associated with the ECPs developed by Bennett et al.~{\cite{2017BEN, 2018BEN}}, respectively. 
Root mean square deviation (RMSD) of the binding energies from CCSD(T) 
reported by \citet{2014DUB} and \citet{2023NAK} are 0.15 kcal/mol 0.315 kcal/mol, respectively. 
Table~{\ref{tab:ccsdt-dmc-comparison}} shows the binding energies obtained in this study by DMC calculations with the aug-ccpV6Z basis sets (without CP) associated with the ECPs developed by Bennett et al.~{\cite{2017BEN, 2018BEN}}, and those obtained by CCSD(T) in the CBS limit taken from Benchmark Energy and Geometry DataBase (BEGDB)~{\cite{2008REZ_BEGDB}}. 
In this work, we obtained a RMSD of 0.135 kcal/mol, which is very close to the value obtained by Dubeck\'{y}, while $\sim$ 0.2 kcal/mol off from the value reported by Nakano et al. 
As mentioned in the previous section, Figure.~{\ref{fig:BSIE-summary}} (a) indicates that the cc-pVTZ basis set without the CP correction shows non-negligible BSIEs and the augmentation (aug-ccpVTZ) reduces the BSIEs significantly. In fact, we got 0.247(14) kcal/mol and 0.131(14) kcal/mol for RMSD with cc-pVTZ and aug-ccpVTZ basis sets, respectively. The obtained cc-pVTZ value (0.247(14) kcal/mol) is very close to the previous Nakano's report (0.315 kcal/mol), although the treatments of the non-local terms are different (DLA was employed in the previous study, while DTM is employed in the present study). Thus, the RMSD obtained by Nakano et al.~{\cite{2023NAK}} should be a little affected by BSIEs, while the values obtained by Dubeck\'{y} et al.~{\cite{2014DUB}} with the augmented basis sets should already reach the CBS limit. Thus, as the benchmark values for the A24 data set, one should refer to the binding energies obtained by Dubeck\'{y} et al.~{\cite{2014DUB}} or those obtained in this work.

\section{Conclusions}
In this study, we investigated two basis-set related errors, BSIEs and BSSEs, in binding energy calculations by ab initio FN-DMC calculations using the A24 benchmark set. We revealed that BSIE and BSSE are not negligible in DMC calculations when a small basis set, such as cc-pVDZ, is used without the CP correction. Our study implies that, to get binding energies in the CBS limit with GTOs, one should use, at least, a medium-size basis set, such as cc-pVQZ or aug-cc-pVTZ basis set. We found that the CP correction is also helpful in DMC calculations to eliminate BSIEs, as in QC calculations. With the CP correction, one can use a smaller basis, such as cc-pVTZ or aug-cc-pVDZ basis sets.
This work raises awareness of BSSEs and BSIEs in binding energy calculations by DMC, which have not been extensively studied previously. In the future, a more comprehensive study should be intriguing to investigate BSIEs in DMC calculations further, e.g., for benchmark sets containing larger molecular or periodic systems.

\section*{Acknowledgments}
K.N. is grateful for computational resources from the Numerical Materials Simulator at National Institute for Materials Science (NIMS).
K.N. is grateful for computational resources of the supercomputer Fugaku provided by RIKEN through the HPCI System Research Projects (Project ID: hp240033).
K.N. acknowledges financial support from MEXT Leading Initiative for Excellent Young Researchers (Grant No.~JPMXS0320220025) and from Iketani Science and Technology Foundation (Grant No.~0361248-A).
B.X.S. acknowledges support from the EPSRC Doctoral Training Partnership (EP/T517847/1).
D.A. and A.Z. acknowledge support from Leverhulme grant no. RPG-2020-038.
D.A. and A.Z. also acknowledge support from the European Union under the Next generation EU (projects 20222FXZ33 and P2022MC742).
The authors acknowledge the use of the UCL Kathleen High Performance Computing Facility (Kathleen@UCL), and associated support services, in the completion of this work. 
This research used resources of the Oak Ridge Leadership Computing Facility at the Oak Ridge National Laboratory, which is supported by the Office of Science of the U.S. Department of Energy under Contract No. DE-AC05-00OR22725). Calculations were also performed using the Cambridge Service for Data Driven Discovery (CSD3) operated by the University of Cambridge Research Computing Service (www.csd3.cam.ac.uk), provided by Dell EMC and Intel using Tier-2 funding from the Engineering and Physical Sciences Research Council (capital grant EP/T022159/1 and EP/P020259/1). This work also used the ARCHER UK National Supercomputing Service (https://www.archer2.ac.uk), the United Kingdom Car Parrinello (UKCP) consortium (EP/F036884/1).

\section*{Code availability} \label{sec:code}
The QMC kernels used in this work, \turborvb\ and \qmcpack, are available from their GitHub repositories, [\url{https://github.com/sissaschool/turborvb}] and [\url{https://github.com/QMCPACK/qmcpack}], respectively.

\section*{Supporting Information Available} \label{sec:code}
Supporting Information includes figures plotting binding energies obtained by the methods employed in this study (HF, MP2, and DMC) with various basis sets.

\bibliography{./references.bib}

\providecommand{\latin}[1]{#1}
\makeatletter
\providecommand{\doi}
  {\begingroup\let\do\@makeother\dospecials
  \catcode`\{=1 \catcode`\}=2 \doi@aux}
\providecommand{\doi@aux}[1]{\endgroup\texttt{#1}}
\makeatother
\providecommand*\mcitethebibliography{\thebibliography}
\csname @ifundefined\endcsname{endmcitethebibliography}  {\let\endmcitethebibliography\endthebibliography}{}
\begin{mcitethebibliography}{63}
\providecommand*\natexlab[1]{#1}
\providecommand*\mciteSetBstSublistMode[1]{}
\providecommand*\mciteSetBstMaxWidthForm[2]{}
\providecommand*\mciteBstWouldAddEndPuncttrue
  {\def\EndOfBibitem{\unskip.}}
\providecommand*\mciteBstWouldAddEndPunctfalse
  {\let\EndOfBibitem\relax}
\providecommand*\mciteSetBstMidEndSepPunct[3]{}
\providecommand*\mciteSetBstSublistLabelBeginEnd[3]{}
\providecommand*\EndOfBibitem{}
\mciteSetBstSublistMode{f}
\mciteSetBstMaxWidthForm{subitem}{(\alph{mcitesubitemcount})}
\mciteSetBstSublistLabelBeginEnd
  {\mcitemaxwidthsubitemform\space}
  {\relax}
  {\relax}

\bibitem[Ceperley(1986)]{1986CEP_GFMC}
Ceperley,~D. The statistical error of green's function Monte Carlo. \emph{J. Stat. Phys.} \textbf{1986}, \emph{43}, 815--826\relax
\mciteBstWouldAddEndPuncttrue
\mciteSetBstMidEndSepPunct{\mcitedefaultmidpunct}
{\mcitedefaultendpunct}{\mcitedefaultseppunct}\relax
\EndOfBibitem
\bibitem[Foulkes \latin{et~al.}(2001)Foulkes, Mitas, Needs, and Rajagopal]{2001FOU_qmcreview}
Foulkes,~W. M.~C.; Mitas,~L.; Needs,~R.~J.; Rajagopal,~G. Quantum Monte Carlo simulations of solids. \emph{Rev. Mod. Phys.} \textbf{2001}, \emph{73}, 33\relax
\mciteBstWouldAddEndPuncttrue
\mciteSetBstMidEndSepPunct{\mcitedefaultmidpunct}
{\mcitedefaultendpunct}{\mcitedefaultseppunct}\relax
\EndOfBibitem
\bibitem[Drummond \latin{et~al.}(2015)Drummond, Monserrat, Lloyd-Williams, R{\'\i}os, Pickard, and Needs]{2015DRU_hydrogen}
Drummond,~N.~D.; Monserrat,~B.; Lloyd-Williams,~J.~H.; R{\'\i}os,~P.~L.; Pickard,~C.~J.; Needs,~R.~J. Quantum Monte Carlo study of the phase diagram of solid molecular hydrogen at extreme pressures. \emph{Nat. Commun.} \textbf{2015}, \emph{6}, 1--6\relax
\mciteBstWouldAddEndPuncttrue
\mciteSetBstMidEndSepPunct{\mcitedefaultmidpunct}
{\mcitedefaultendpunct}{\mcitedefaultseppunct}\relax
\EndOfBibitem
\bibitem[Mazzola \latin{et~al.}(2018)Mazzola, Helled, and Sorella]{2018MAZ_hydrogen}
Mazzola,~G.; Helled,~R.; Sorella,~S. Phase diagram of hydrogen and a hydrogen-helium mixture at planetary conditions by Quantum Monte Carlo simulations. \emph{Phys. Rev. Lett.} \textbf{2018}, \emph{120}, 025701\relax
\mciteBstWouldAddEndPuncttrue
\mciteSetBstMidEndSepPunct{\mcitedefaultmidpunct}
{\mcitedefaultendpunct}{\mcitedefaultseppunct}\relax
\EndOfBibitem
\bibitem[Tirelli \latin{et~al.}(2022)Tirelli, Tenti, Nakano, and Sorella]{2022TIR_hydrogen_MLP}
Tirelli,~A.; Tenti,~G.; Nakano,~K.; Sorella,~S. High-pressure hydrogen by machine learning and quantum Monte Carlo. \emph{Phys. Rev. B} \textbf{2022}, \emph{106}, L041105\relax
\mciteBstWouldAddEndPuncttrue
\mciteSetBstMidEndSepPunct{\mcitedefaultmidpunct}
{\mcitedefaultendpunct}{\mcitedefaultseppunct}\relax
\EndOfBibitem
\bibitem[Ly and Ceperley(2022)Ly, and Ceperley]{2022_LY_hydrogen}
Ly,~K.~K.; Ceperley,~D.~M. {Phonons of metallic hydrogen with quantum Monte Carlo}. \emph{J. Chem. Phys.} \textbf{2022}, \emph{156}, 044108\relax
\mciteBstWouldAddEndPuncttrue
\mciteSetBstMidEndSepPunct{\mcitedefaultmidpunct}
{\mcitedefaultendpunct}{\mcitedefaultseppunct}\relax
\EndOfBibitem
\bibitem[Niu \latin{et~al.}(2023)Niu, Yang, Jensen, Holzmann, Pierleoni, and Ceperley]{2023NIU_hyrdogen_MLP}
Niu,~H.; Yang,~Y.; Jensen,~S.; Holzmann,~M.; Pierleoni,~C.; Ceperley,~D.~M. Stable Solid Molecular Hydrogen above 900 K from a Machine-Learned Potential Trained with Diffusion Quantum Monte Carlo. \emph{Phys. Rev. Lett.} \textbf{2023}, \emph{130}, 076102\relax
\mciteBstWouldAddEndPuncttrue
\mciteSetBstMidEndSepPunct{\mcitedefaultmidpunct}
{\mcitedefaultendpunct}{\mcitedefaultseppunct}\relax
\EndOfBibitem
\bibitem[Monacelli \latin{et~al.}(2023)Monacelli, Casula, Nakano, Sorella, and Mauri]{2023MON_hydrogen}
Monacelli,~L.; Casula,~M.; Nakano,~K.; Sorella,~S.; Mauri,~F. Quantum phase diagram of high-pressure hydrogen. \emph{Nat. Phys.} \textbf{2023}, \emph{19}, 845--850\relax
\mciteBstWouldAddEndPuncttrue
\mciteSetBstMidEndSepPunct{\mcitedefaultmidpunct}
{\mcitedefaultendpunct}{\mcitedefaultseppunct}\relax
\EndOfBibitem
\bibitem[Tenti \latin{et~al.}(2024)Tenti, Nakano, Tirelli, Sorella, and Casula]{2024GIA_hydrogen_MLP}
Tenti,~G.; Nakano,~K.; Tirelli,~A.; Sorella,~S.; Casula,~M. Principal deuterium Hugoniot via quantum Monte Carlo and $\mathrm{\ensuremath{\Delta}}$-learning. \emph{Phys. Rev. B} \textbf{2024}, \emph{110}, L041107\relax
\mciteBstWouldAddEndPuncttrue
\mciteSetBstMidEndSepPunct{\mcitedefaultmidpunct}
{\mcitedefaultendpunct}{\mcitedefaultseppunct}\relax
\EndOfBibitem
\bibitem[Krogel \latin{et~al.}(2020)Krogel, Yuk, Kent, and Cooper]{2020KRO_TiS2}
Krogel,~J.~T.; Yuk,~S.~F.; Kent,~P. R.~C.; Cooper,~V.~R. Perspectives on van der Waals Density Functionals: The Case of TiS$_2$. \emph{J. Phys. Chem. A} \textbf{2020}, \emph{124}, 9867--9876\relax
\mciteBstWouldAddEndPuncttrue
\mciteSetBstMidEndSepPunct{\mcitedefaultmidpunct}
{\mcitedefaultendpunct}{\mcitedefaultseppunct}\relax
\EndOfBibitem
\bibitem[Ichibha \latin{et~al.}(2021)Ichibha, Dzubak, Krogel, Cooper, and Reboredo]{2021ICH_CrI3}
Ichibha,~T.; Dzubak,~A.~L.; Krogel,~J.~T.; Cooper,~V.~R.; Reboredo,~F.~A. ${\mathrm{CrI}}_{3}$ revisited with a many-body ab initio theoretical approach. \emph{Phys. Rev. Mater.} \textbf{2021}, \emph{5}, 064006\relax
\mciteBstWouldAddEndPuncttrue
\mciteSetBstMidEndSepPunct{\mcitedefaultmidpunct}
{\mcitedefaultendpunct}{\mcitedefaultseppunct}\relax
\EndOfBibitem
\bibitem[Nikaido \latin{et~al.}(2022)Nikaido, Ichibha, Hongo, Reboredo, Kumar, Mahadevan, Maezono, and Nakano]{2022NIK_BNs}
Nikaido,~Y.; Ichibha,~T.; Hongo,~K.; Reboredo,~F.~A.; Kumar,~K. C.~H.; Mahadevan,~P.; Maezono,~R.; Nakano,~K. Diffusion Monte Carlo Study on Relative Stabilities of Boron Nitride Polymorphs. \emph{J. Phys. Chem. C} \textbf{2022}, \emph{126}, 6000--6007\relax
\mciteBstWouldAddEndPuncttrue
\mciteSetBstMidEndSepPunct{\mcitedefaultmidpunct}
{\mcitedefaultendpunct}{\mcitedefaultseppunct}\relax
\EndOfBibitem
\bibitem[Wines \latin{et~al.}(2023)Wines, Choudhary, and Tavazza]{2023WIN_CrX3}
Wines,~D.; Choudhary,~K.; Tavazza,~F. Systematic DFT+U and Quantum Monte Carlo Benchmark of Magnetic Two-Dimensional (2D) CrX$_3$ (X = I, Br, Cl, F). \emph{J. Phys. Chem. C} \textbf{2023}, \emph{127}, 1176--1188\relax
\mciteBstWouldAddEndPuncttrue
\mciteSetBstMidEndSepPunct{\mcitedefaultmidpunct}
{\mcitedefaultendpunct}{\mcitedefaultseppunct}\relax
\EndOfBibitem
\bibitem[Wines \latin{et~al.}(2023)Wines, Tiihonen, Saritas, Krogel, and Ataca]{2023WIN_VSe2}
Wines,~D.; Tiihonen,~J.; Saritas,~K.; Krogel,~J.~T.; Ataca,~C. A Quantum Monte Carlo Study of the Structural, Energetic, and Magnetic Properties of Two-Dimensional H and T Phase VSe$_2$. \emph{J. Phys. Chem. Lett.} \textbf{2023}, \emph{14}, 3553--3560\relax
\mciteBstWouldAddEndPuncttrue
\mciteSetBstMidEndSepPunct{\mcitedefaultmidpunct}
{\mcitedefaultendpunct}{\mcitedefaultseppunct}\relax
\EndOfBibitem
\bibitem[Zen \latin{et~al.}(2018)Zen, Brandenburg, Klime{\v{s}}, Tkatchenko, Alf{\`e}, and Michaelides]{2018ZEN_MolCrys}
Zen,~A.; Brandenburg,~J.~G.; Klime{\v{s}},~J.; Tkatchenko,~A.; Alf{\`e},~D.; Michaelides,~A. Fast and accurate quantum Monte Carlo for molecular crystals. \emph{Proc. Natl. Acad. Sci. U.S.A.} \textbf{2018}, \emph{115}, 1724--1729\relax
\mciteBstWouldAddEndPuncttrue
\mciteSetBstMidEndSepPunct{\mcitedefaultmidpunct}
{\mcitedefaultendpunct}{\mcitedefaultseppunct}\relax
\EndOfBibitem
\bibitem[Della~Pia \latin{et~al.}(2024)Della~Pia, Zen, Alf\`e, and Michaelides]{2024FLA_DMC_MolCrys}
Della~Pia,~F.; Zen,~A.; Alf\`e,~D.; Michaelides,~A. How Accurate Are Simulations and Experiments for the Lattice Energies of Molecular Crystals? \emph{Phys. Rev. Lett.} \textbf{2024}, \emph{133}, 046401\relax
\mciteBstWouldAddEndPuncttrue
\mciteSetBstMidEndSepPunct{\mcitedefaultmidpunct}
{\mcitedefaultendpunct}{\mcitedefaultseppunct}\relax
\EndOfBibitem
\bibitem[Beaudet \latin{et~al.}(2008)Beaudet, Casula, Kim, Sorella, and Martin]{2008BEA_absorption}
Beaudet,~T.~D.; Casula,~M.; Kim,~J.; Sorella,~S.; Martin,~R.~M. Molecular hydrogen adsorbed on benzene: Insights from a quantum Monte Carlo study. \emph{J. Chem. Phys.} \textbf{2008}, \emph{129}, 164711\relax
\mciteBstWouldAddEndPuncttrue
\mciteSetBstMidEndSepPunct{\mcitedefaultmidpunct}
{\mcitedefaultendpunct}{\mcitedefaultseppunct}\relax
\EndOfBibitem
\bibitem[Zen \latin{et~al.}(2016)Zen, Roch, Cox, Hu, Sorella, Alf\`{e}, and Michaelides]{2016ZEN_absorption}
Zen,~A.; Roch,~L.~M.; Cox,~S.~J.; Hu,~X.~L.; Sorella,~S.; Alf\`{e},~D.; Michaelides,~A. Toward accurate adsorption energetics on clay surfaces. \emph{J. Phys. Chem. C} \textbf{2016}, \emph{120}, 26402--26413\relax
\mciteBstWouldAddEndPuncttrue
\mciteSetBstMidEndSepPunct{\mcitedefaultmidpunct}
{\mcitedefaultendpunct}{\mcitedefaultseppunct}\relax
\EndOfBibitem
\bibitem[Al-Hamdani \latin{et~al.}(2017)Al-Hamdani, Rossi, Alfè, Tsatsoulis, Ramberger, Brandenburg, Zen, Kresse, Grüneis, Tkatchenko, and Michaelides]{2017HAM_absorption}
Al-Hamdani,~Y.~S.; Rossi,~M.; Alfè,~D.; Tsatsoulis,~T.; Ramberger,~B.; Brandenburg,~J.~G.; Zen,~A.; Kresse,~G.; Grüneis,~A.; Tkatchenko,~A.; Michaelides,~A. {Properties of the water to boron nitride interaction: From zero to two dimensions with benchmark accuracy}. \emph{J. Chem. Phys,} \textbf{2017}, \emph{147}, 044710\relax
\mciteBstWouldAddEndPuncttrue
\mciteSetBstMidEndSepPunct{\mcitedefaultmidpunct}
{\mcitedefaultendpunct}{\mcitedefaultseppunct}\relax
\EndOfBibitem
\bibitem[Hsing \latin{et~al.}(2019)Hsing, Chang, Cheng, and Wei]{2019CHE_absorption}
Hsing,~C.-R.; Chang,~C.-M.; Cheng,~C.; Wei,~C.-M. Quantum Monte Carlo Studies of CO Adsorption on Transition Metal Surfaces. \emph{J. Phys. Chem. C} \textbf{2019}, \emph{123}, 15659--15664\relax
\mciteBstWouldAddEndPuncttrue
\mciteSetBstMidEndSepPunct{\mcitedefaultmidpunct}
{\mcitedefaultendpunct}{\mcitedefaultseppunct}\relax
\EndOfBibitem
\bibitem[Shi \latin{et~al.}(2023)Shi, Zen, Kapil, Nagy, Grüneis, and Michaelides]{2023BEN_absorption}
Shi,~B.~X.; Zen,~A.; Kapil,~V.; Nagy,~P.~R.; Grüneis,~A.; Michaelides,~A. Many-Body Methods for Surface Chemistry Come of Age: Achieving Consensus with Experiments. \emph{J. Am. Chem. Soc.} \textbf{2023}, \emph{145}, 25372--25381\relax
\mciteBstWouldAddEndPuncttrue
\mciteSetBstMidEndSepPunct{\mcitedefaultmidpunct}
{\mcitedefaultendpunct}{\mcitedefaultseppunct}\relax
\EndOfBibitem
\bibitem[Korth \latin{et~al.}(2011)Korth, Grimme, and Towler]{2011KOR}
Korth,~M.; Grimme,~S.; Towler,~M.~D. The Lithium--Thiophene Riddle Revisited. \emph{J. Phys. Chem. A} \textbf{2011}, \emph{115}, 11734--11739\relax
\mciteBstWouldAddEndPuncttrue
\mciteSetBstMidEndSepPunct{\mcitedefaultmidpunct}
{\mcitedefaultendpunct}{\mcitedefaultseppunct}\relax
\EndOfBibitem
\bibitem[Nemec \latin{et~al.}(2010)Nemec, Towler, and Needs]{2010NEM_G2set}
Nemec,~N.; Towler,~M.~D.; Needs,~R.~J. {Benchmark all-electron ab initio quantum Monte Carlo calculations for small molecules}. \emph{J. Chem. Phys.} \textbf{2010}, \emph{132}, 034111\relax
\mciteBstWouldAddEndPuncttrue
\mciteSetBstMidEndSepPunct{\mcitedefaultmidpunct}
{\mcitedefaultendpunct}{\mcitedefaultseppunct}\relax
\EndOfBibitem
\bibitem[Petruzielo \latin{et~al.}(2012)Petruzielo, Toulouse, and Umrigar]{2012JCP_Petruzielo_DMC-G2}
Petruzielo,~F.~R.; Toulouse,~J.; Umrigar,~C.~J. Approaching chemical accuracy with quantum Monte Carlo. \emph{J. Chem. Phys.} \textbf{2012}, \emph{136}, 124116\relax
\mciteBstWouldAddEndPuncttrue
\mciteSetBstMidEndSepPunct{\mcitedefaultmidpunct}
{\mcitedefaultendpunct}{\mcitedefaultseppunct}\relax
\EndOfBibitem
\bibitem[Zen \latin{et~al.}(2013)Zen, Luo, Sorella, and Guidoni]{2013JCTC_Zen_water}
Zen,~A.; Luo,~Y.; Sorella,~S.; Guidoni,~L. Molecular Properties by Quantum Monte Carlo: An Investigation on the Role of the Wave Function Ansatz and the Basis Set in the Water Molecule. \emph{J. Chem. Theory Comput.} \textbf{2013}, \emph{9}, 4332--4350\relax
\mciteBstWouldAddEndPuncttrue
\mciteSetBstMidEndSepPunct{\mcitedefaultmidpunct}
{\mcitedefaultendpunct}{\mcitedefaultseppunct}\relax
\EndOfBibitem
\bibitem[Scemama \latin{et~al.}(2016)Scemama, Applencourt, Giner, and Caffarel]{Scemama2016}
Scemama,~A.; Applencourt,~T.; Giner,~E.; Caffarel,~M. {Quantum Monte Carlo with very large multideterminant wavefunctions}. \emph{J. Comput. Chem.} \textbf{2016}, \emph{37}, 1866--1875\relax
\mciteBstWouldAddEndPuncttrue
\mciteSetBstMidEndSepPunct{\mcitedefaultmidpunct}
{\mcitedefaultendpunct}{\mcitedefaultseppunct}\relax
\EndOfBibitem
\bibitem[Caffarel \latin{et~al.}(2016)Caffarel, Applencourt, Giner, and Scemama]{2016MIC}
Caffarel,~M.; Applencourt,~T.; Giner,~E.; Scemama,~A. Communication: Toward an improved control of the fixed-node error in quantum Monte Carlo: The case of the water molecule. \emph{J. Chem. Phys.} \textbf{2016}, \emph{144}, 151103\relax
\mciteBstWouldAddEndPuncttrue
\mciteSetBstMidEndSepPunct{\mcitedefaultmidpunct}
{\mcitedefaultendpunct}{\mcitedefaultseppunct}\relax
\EndOfBibitem
\bibitem[Scemama \latin{et~al.}(2020)Scemama, Giner, Benali, and Loos]{2020JCP_Scemama_FNDMC-rangesep}
Scemama,~A.; Giner,~E.; Benali,~A.; Loos,~P.-F. Taming the fixed-node error in diffusion Monte Carlo via range separation. \emph{J. Chem. Phys.} \textbf{2020}, \emph{153}, 174107\relax
\mciteBstWouldAddEndPuncttrue
\mciteSetBstMidEndSepPunct{\mcitedefaultmidpunct}
{\mcitedefaultendpunct}{\mcitedefaultseppunct}\relax
\EndOfBibitem
\bibitem[Korth \latin{et~al.}(2008)Korth, L{\"u}chow, and Grimme]{2008KOR_S22set}
Korth,~M.; L{\"u}chow,~A.; Grimme,~S. Toward the exact solution of the electronic Schr{\"o}dinger equation for noncovalent molecular interactions: worldwide distributed quantum Monte Carlo calculations. \emph{J. Phys. Chem. A} \textbf{2008}, \emph{112}, 2104--2109\relax
\mciteBstWouldAddEndPuncttrue
\mciteSetBstMidEndSepPunct{\mcitedefaultmidpunct}
{\mcitedefaultendpunct}{\mcitedefaultseppunct}\relax
\EndOfBibitem
\bibitem[Dubecky \latin{et~al.}(2013)Dubecky, Jurecka, Derian, Hobza, Otyepka, and Mitas]{2013DUB}
Dubecky,~M.; Jurecka,~P.; Derian,~R.; Hobza,~P.; Otyepka,~M.; Mitas,~L. Quantum Monte Carlo methods describe noncovalent interactions with subchemical accuracy. \emph{J. Chem. Theory Comput.} \textbf{2013}, \emph{9}, 4287--4292\relax
\mciteBstWouldAddEndPuncttrue
\mciteSetBstMidEndSepPunct{\mcitedefaultmidpunct}
{\mcitedefaultendpunct}{\mcitedefaultseppunct}\relax
\EndOfBibitem
\bibitem[Dubeck{\`y} \latin{et~al.}(2014)Dubeck{\`y}, Derian, Jure{\v{c}}ka, Mitas, Hobza, and Otyepka]{2014DUB}
Dubeck{\`y},~M.; Derian,~R.; Jure{\v{c}}ka,~P.; Mitas,~L.; Hobza,~P.; Otyepka,~M. Quantum Monte Carlo for noncovalent interactions: an efficient protocol attaining benchmark accuracy. \emph{Phys. Chem. Chem. Phys.} \textbf{2014}, \emph{16}, 20915--20923\relax
\mciteBstWouldAddEndPuncttrue
\mciteSetBstMidEndSepPunct{\mcitedefaultmidpunct}
{\mcitedefaultendpunct}{\mcitedefaultseppunct}\relax
\EndOfBibitem
\bibitem[Mostaani \latin{et~al.}(2015)Mostaani, Drummond, and Fal'ko]{2015PRL_bilayergraphene}
Mostaani,~E.; Drummond,~N.~D.; Fal'ko,~V.~I. Quantum Monte Carlo Calculation of the Binding Energy of Bilayer Graphene. \emph{Phys. Rev. Lett.} \textbf{2015}, \emph{115}, 115501\relax
\mciteBstWouldAddEndPuncttrue
\mciteSetBstMidEndSepPunct{\mcitedefaultmidpunct}
{\mcitedefaultendpunct}{\mcitedefaultseppunct}\relax
\EndOfBibitem
\bibitem[Nakano \latin{et~al.}(2019)Nakano, Maezono, and Sorella]{2019NAK_vdw}
Nakano,~K.; Maezono,~R.; Sorella,~S. {All-Electron Quantum Monte Carlo with Jastrow Single Determinant Ansatz: Application to the Sodium Dimer}. \emph{J. Chem. Theory Comput.} \textbf{2019}, \emph{15}, 4044--4055\relax
\mciteBstWouldAddEndPuncttrue
\mciteSetBstMidEndSepPunct{\mcitedefaultmidpunct}
{\mcitedefaultendpunct}{\mcitedefaultseppunct}\relax
\EndOfBibitem
\bibitem[Zen \latin{et~al.}(2018)Zen, Brandenburg, Klime{\v{s}}, Tkatchenko, Alf{\`{e}}, and Michaelides]{Zen2018_PNAS}
Zen,~A.; Brandenburg,~J.~G.; Klime{\v{s}},~J.; Tkatchenko,~A.; Alf{\`{e}},~D.; Michaelides,~A. {Fast and accurate quantum Monte Carlo for molecular crystals}. \emph{Proc. Natl. Acad. Sci. U.S.A.} \textbf{2018}, \emph{115}, 1724--1729\relax
\mciteBstWouldAddEndPuncttrue
\mciteSetBstMidEndSepPunct{\mcitedefaultmidpunct}
{\mcitedefaultendpunct}{\mcitedefaultseppunct}\relax
\EndOfBibitem
\bibitem[Benali \latin{et~al.}(2020)Benali, Shin, and Heinonen]{2020BEN_L7set}
Benali,~A.; Shin,~H.; Heinonen,~O. {Quantum Monte Carlo benchmarking of large noncovalent complexes in the L7 benchmark set}. \emph{J. Chem. Phys.} \textbf{2020}, \emph{153}, 194113\relax
\mciteBstWouldAddEndPuncttrue
\mciteSetBstMidEndSepPunct{\mcitedefaultmidpunct}
{\mcitedefaultendpunct}{\mcitedefaultseppunct}\relax
\EndOfBibitem
\bibitem[Al-Hamdani \latin{et~al.}(2021)Al-Hamdani, Nagy, Zen, Barton, Kállay, Brandenburg, and Tkatchenko]{2021YAS_C60_CPPA}
Al-Hamdani,~Y.~S.; Nagy,~P.~R.; Zen,~A.; Barton,~D.; Kállay,~M.; Brandenburg,~J.~G.; Tkatchenko,~A. Interactions between large molecules pose a puzzle for reference quantum mechanical methods. \emph{Nat. Commun.} \textbf{2021}, \emph{12}, 3927\relax
\mciteBstWouldAddEndPuncttrue
\mciteSetBstMidEndSepPunct{\mcitedefaultmidpunct}
{\mcitedefaultendpunct}{\mcitedefaultseppunct}\relax
\EndOfBibitem
\bibitem[Raghav \latin{et~al.}(2023)Raghav, Maezono, Hongo, Sorella, and Nakano]{2023RAG_G2set}
Raghav,~A.; Maezono,~R.; Hongo,~K.; Sorella,~S.; Nakano,~K. \emph{J. Chem. Theory Comput.} \textbf{2023}, \emph{19}, 2222--2229\relax
\mciteBstWouldAddEndPuncttrue
\mciteSetBstMidEndSepPunct{\mcitedefaultmidpunct}
{\mcitedefaultendpunct}{\mcitedefaultseppunct}\relax
\EndOfBibitem
\bibitem[Dubeck{\'{y}} \latin{et~al.}(2016)Dubeck{\'{y}}, Mitas, and Jure{\v{c}}ka]{Dubecky2016}
Dubeck{\'{y}},~M.; Mitas,~L.; Jure{\v{c}}ka,~P. {Noncovalent Interactions by Quantum Monte Carlo}. \emph{Chem. Rev.} \textbf{2016}, \emph{116}, 5188--5215\relax
\mciteBstWouldAddEndPuncttrue
\mciteSetBstMidEndSepPunct{\mcitedefaultmidpunct}
{\mcitedefaultendpunct}{\mcitedefaultseppunct}\relax
\EndOfBibitem
\bibitem[Nakano \latin{et~al.}(2024)Nakano, Sorella, Alfè, and Zen]{2024NAK_JAGPn}
Nakano,~K.; Sorella,~S.; Alfè,~D.; Zen,~A. Beyond Single-Reference Fixed-Node Approximation in Ab Initio Diffusion Monte Carlo Using Antisymmetrized Geminal Power Applied to Systems with Hundreds of Electrons. \emph{J. Chem. Theory Comput.} \textbf{2024}, \emph{20}, 4591--4604\relax
\mciteBstWouldAddEndPuncttrue
\mciteSetBstMidEndSepPunct{\mcitedefaultmidpunct}
{\mcitedefaultendpunct}{\mcitedefaultseppunct}\relax
\EndOfBibitem
\bibitem[Sch{\"a}fer \latin{et~al.}(2024)Sch{\"a}fer, Irmler, Gallo, and Gr{\"u}neis]{2024SCH_CCSDTQ}
Sch{\"a}fer,~T.; Irmler,~A.; Gallo,~A.; Gr{\"u}neis,~A. Understanding discrepancies of wavefunction theories for large molecules. \emph{arXiv preprint arXiv:2407.01442} \textbf{2024}, \relax
\mciteBstWouldAddEndPunctfalse
\mciteSetBstMidEndSepPunct{\mcitedefaultmidpunct}
{}{\mcitedefaultseppunct}\relax
\EndOfBibitem
\bibitem[Van~Duijneveldt \latin{et~al.}(1994)Van~Duijneveldt, van Duijneveldt-van~de Rijdt, and van Lenthe]{1994DUI}
Van~Duijneveldt,~F.~B.; van Duijneveldt-van~de Rijdt,~J.~G.; van Lenthe,~J.~H. State of the art in counterpoise theory. \emph{Chem. Rev.} \textbf{1994}, \emph{94}, 1873--1885\relax
\mciteBstWouldAddEndPuncttrue
\mciteSetBstMidEndSepPunct{\mcitedefaultmidpunct}
{\mcitedefaultendpunct}{\mcitedefaultseppunct}\relax
\EndOfBibitem
\bibitem[Dunning(2000)]{2000DUN}
Dunning,~T.~H. A road map for the calculation of molecular binding energies. \emph{J. Phys. Chem. A} \textbf{2000}, \emph{104}, 9062--9080\relax
\mciteBstWouldAddEndPuncttrue
\mciteSetBstMidEndSepPunct{\mcitedefaultmidpunct}
{\mcitedefaultendpunct}{\mcitedefaultseppunct}\relax
\EndOfBibitem
\bibitem[Dunning~Jr(1989)]{1989DUN}
Dunning~Jr,~T.~H. Gaussian basis sets for use in correlated molecular calculations. I. The atoms boron through neon and hydrogen. \emph{J. Chem. Phys.} \textbf{1989}, \emph{90}, 1007--1023\relax
\mciteBstWouldAddEndPuncttrue
\mciteSetBstMidEndSepPunct{\mcitedefaultmidpunct}
{\mcitedefaultendpunct}{\mcitedefaultseppunct}\relax
\EndOfBibitem
\bibitem[Boys and Bernardi(1970)Boys, and Bernardi]{1970BOY}
Boys,~S.; Bernardi,~F. The calculation of small molecular interactions by the differences of separate total energies. Some procedures with reduced errors. \emph{Mol. Phys.} \textbf{1970}, \emph{19}, 553--566\relax
\mciteBstWouldAddEndPuncttrue
\mciteSetBstMidEndSepPunct{\mcitedefaultmidpunct}
{\mcitedefaultendpunct}{\mcitedefaultseppunct}\relax
\EndOfBibitem
\bibitem[Zhou \latin{et~al.}(2024)Zhou, Huang, and He]{2024JCP_Zhou_WatClusters}
Zhou,~X.; Huang,~Z.; He,~X. Diffusion Monte Carlo method for barrier heights of multiple proton exchanges and complexation energies in small water, ammonia, and hydrogen fluoride clusters. \emph{J. Chem. Phys.} \textbf{2024}, \emph{160}, 054103\relax
\mciteBstWouldAddEndPuncttrue
\mciteSetBstMidEndSepPunct{\mcitedefaultmidpunct}
{\mcitedefaultendpunct}{\mcitedefaultseppunct}\relax
\EndOfBibitem
\bibitem[\v{R}ez\'a\v{c} and Hobza(2013)\v{R}ez\'a\v{c}, and Hobza]{2013REZ}
\v{R}ez\'a\v{c},~J.; Hobza,~P. Describing noncovalent interactions beyond the common approximations: how accurate is the “gold standard,” CCSD (T) at the complete basis set limit? \emph{J. Chem. Theory Comput.} \textbf{2013}, \emph{9}, 2151--2155\relax
\mciteBstWouldAddEndPuncttrue
\mciteSetBstMidEndSepPunct{\mcitedefaultmidpunct}
{\mcitedefaultendpunct}{\mcitedefaultseppunct}\relax
\EndOfBibitem
\bibitem[Bennett \latin{et~al.}(2017)Bennett, Melton, Annaberdiyev, Wang, Shulenburger, and Mitas]{2017BEN}
Bennett,~M.~C.; Melton,~C.~A.; Annaberdiyev,~A.; Wang,~G.; Shulenburger,~L.; Mitas,~L. {A new generation of effective core potentials for correlated calculations}. \emph{J. Chem. Phys} \textbf{2017}, \emph{147}, 224106\relax
\mciteBstWouldAddEndPuncttrue
\mciteSetBstMidEndSepPunct{\mcitedefaultmidpunct}
{\mcitedefaultendpunct}{\mcitedefaultseppunct}\relax
\EndOfBibitem
\bibitem[Bennett \latin{et~al.}(2018)Bennett, Wang, Annaberdiyev, Melton, Shulenburger, and Mitas]{2018BEN}
Bennett,~M.~C.; Wang,~G.; Annaberdiyev,~A.; Melton,~C.~A.; Shulenburger,~L.; Mitas,~L. {A new generation of effective core potentials from correlated calculations: 2nd row elements}. \emph{J. Chem. Phys.} \textbf{2018}, \emph{149}, 104108\relax
\mciteBstWouldAddEndPuncttrue
\mciteSetBstMidEndSepPunct{\mcitedefaultmidpunct}
{\mcitedefaultendpunct}{\mcitedefaultseppunct}\relax
\EndOfBibitem
\bibitem[Nakano \latin{et~al.}(2020)Nakano, Attaccalite, Barborini, Capriotti, Casula, Coccia, Dagrada, Genovese, Luo, Mazzola, Zen, and Sorella]{2020NAK_turborvb}
Nakano,~K.; Attaccalite,~C.; Barborini,~M.; Capriotti,~L.; Casula,~M.; Coccia,~E.; Dagrada,~M.; Genovese,~C.; Luo,~Y.; Mazzola,~G.; Zen,~A.; Sorella,~S. {TurboRVB: A many-body toolkit for ab initio electronic simulations by quantum Monte Carlo}. \emph{J. Chem. Phys.} \textbf{2020}, \emph{152}, 204121\relax
\mciteBstWouldAddEndPuncttrue
\mciteSetBstMidEndSepPunct{\mcitedefaultmidpunct}
{\mcitedefaultendpunct}{\mcitedefaultseppunct}\relax
\EndOfBibitem
\bibitem[Casula \latin{et~al.}(2005)Casula, Filippi, and Sorella]{2005CAS_lrdmc}
Casula,~M.; Filippi,~C.; Sorella,~S. Diffusion Monte Carlo method with lattice regularization. \emph{Phys. Rev. Lett.} \textbf{2005}, \emph{95}, 100201\relax
\mciteBstWouldAddEndPuncttrue
\mciteSetBstMidEndSepPunct{\mcitedefaultmidpunct}
{\mcitedefaultendpunct}{\mcitedefaultseppunct}\relax
\EndOfBibitem
\bibitem[Zen \latin{et~al.}(2019)Zen, Brandenburg, Michaelides, and Alfè]{2019ZEN}
Zen,~A.; Brandenburg,~J.~G.; Michaelides,~A.; Alfè,~D. A new scheme for fixed node diffusion quantum Monte Carlo with pseudopotentials: Improving reproducibility and reducing the trial-wave-function bias. \emph{J. Chem. Phys.} \textbf{2019}, \emph{151}, 134105\relax
\mciteBstWouldAddEndPuncttrue
\mciteSetBstMidEndSepPunct{\mcitedefaultmidpunct}
{\mcitedefaultendpunct}{\mcitedefaultseppunct}\relax
\EndOfBibitem
\bibitem[Posenitskiy \latin{et~al.}(2023)Posenitskiy, Chilkuri, Ammar, Hapka, Pernal, Shinde, Landinez~Borda, Filippi, Nakano, Kohulák, Sorella, de~Oliveira~Castro, Jalby, Ríos, Alavi, and Scemama]{2023POS}
Posenitskiy,~E.; Chilkuri,~V.~G.; Ammar,~A.; Hapka,~M.; Pernal,~K.; Shinde,~R.; Landinez~Borda,~E.~J.; Filippi,~C.; Nakano,~K.; Kohulák,~O.; Sorella,~S.; de~Oliveira~Castro,~P.; Jalby,~W.; Ríos,~P.~L.; Alavi,~A.; Scemama,~A. {TREXIO: A file format and library for quantum chemistry}. \emph{J. Chem. Phys.} \textbf{2023}, \emph{158}, 174801\relax
\mciteBstWouldAddEndPuncttrue
\mciteSetBstMidEndSepPunct{\mcitedefaultmidpunct}
{\mcitedefaultendpunct}{\mcitedefaultseppunct}\relax
\EndOfBibitem
\bibitem[Sun \latin{et~al.}(2018)Sun, Berkelbach, Blunt, Booth, Guo, Li, Liu, McClain, Sayfutyarova, Sharma, \latin{et~al.} others]{2018SUN}
Sun,~Q.; Berkelbach,~T.~C.; Blunt,~N.~S.; Booth,~G.~H.; Guo,~S.; Li,~Z.; Liu,~J.; McClain,~J.~D.; Sayfutyarova,~E.~R.; Sharma,~S.; others PySCF: the Python-based simulations of chemistry framework. \emph{Wiley Interdiscip. Rev. Comput. Mol. Sci.} \textbf{2018}, \emph{8}, e1340\relax
\mciteBstWouldAddEndPuncttrue
\mciteSetBstMidEndSepPunct{\mcitedefaultmidpunct}
{\mcitedefaultendpunct}{\mcitedefaultseppunct}\relax
\EndOfBibitem
\bibitem[Sun \latin{et~al.}(2020)Sun, Zhang, Banerjee, Bao, Barbry, Blunt, Bogdanov, Booth, Chen, Cui, Eriksen, Gao, Guo, Hermann, Hermes, Koh, Koval, Lehtola, Li, Liu, Mardirossian, McClain, Motta, Mussard, Pham, Pulkin, Purwanto, Robinson, Ronca, Sayfutyarova, Scheurer, Schurkus, Smith, Sun, Sun, Upadhyay, Wagner, Wang, White, Whitfield, Williamson, Wouters, Yang, Yu, Zhu, Berkelbach, Sharma, Sokolov, and Chan]{2020SUN}
Sun,~Q.; Zhang,~X.; Banerjee,~S.; Bao,~P.; Barbry,~M.; Blunt,~N.~S.; Bogdanov,~N.~A.; Booth,~G.~H.; Chen,~J.; Cui,~Z.~H.; Eriksen,~J.~J.; Gao,~Y.; Guo,~S.; Hermann,~J.; Hermes,~M.~R.; Koh,~K.; Koval,~P.; Lehtola,~S.; Li,~Z.; Liu,~J.; Mardirossian,~N.; McClain,~J.~D.; Motta,~M.; Mussard,~B.; Pham,~H.~Q.; Pulkin,~A.; Purwanto,~W.; Robinson,~P.~J.; Ronca,~E.; Sayfutyarova,~E.~R.; Scheurer,~M.; Schurkus,~H.~F.; Smith,~J.~E.; Sun,~C.; Sun,~S.~N.; Upadhyay,~S.; Wagner,~L.~K.; Wang,~X.; White,~A.; Whitfield,~J.~D.; Williamson,~M.~J.; Wouters,~S.; Yang,~J.; Yu,~J.~M.; Zhu,~T.; Berkelbach,~T.~C.; Sharma,~S.; Sokolov,~A.~Y.; Chan,~G. K.~L. Recent developments in the PySCF program package. \emph{J. Chem. Phys.} \textbf{2020}, \emph{153}, 024109\relax
\mciteBstWouldAddEndPuncttrue
\mciteSetBstMidEndSepPunct{\mcitedefaultmidpunct}
{\mcitedefaultendpunct}{\mcitedefaultseppunct}\relax
\EndOfBibitem
\bibitem[Perdew and Zunger(1981)Perdew, and Zunger]{1981PER}
Perdew,~J.~P.; Zunger,~A. Self-interaction correction to density-functional approximations for many-electron systems. \emph{Phys. Rev. B} \textbf{1981}, \emph{23}, 5048\relax
\mciteBstWouldAddEndPuncttrue
\mciteSetBstMidEndSepPunct{\mcitedefaultmidpunct}
{\mcitedefaultendpunct}{\mcitedefaultseppunct}\relax
\EndOfBibitem
\bibitem[Sorella(1998)]{1998SOR}
Sorella,~S. Green function Monte Carlo with stochastic reconfiguration. \emph{Phys. Rev. Lett.} \textbf{1998}, \emph{80}, 4558\relax
\mciteBstWouldAddEndPuncttrue
\mciteSetBstMidEndSepPunct{\mcitedefaultmidpunct}
{\mcitedefaultendpunct}{\mcitedefaultseppunct}\relax
\EndOfBibitem
\bibitem[Kim \latin{et~al.}(2018)Kim, Baczewski, Beaudet, Benali, Bennett, Berrill, Blunt, Borda, Casula, Ceperley, Chiesa, Clark, Clay, Delaney, Dewing, Esler, Hao, Heinonen, Kent, Krogel, Kyl{\"a}np{\"a}{\"a}, Li, Lopez, Luo, Malone, Martin, Mathuriya, McMinis, Melton, Mitas, Morales, Neuscamman, Parker, Flores, Romero, Rubenstein, Shea, Shin, Shulenburger, Tillack, Townsend, Tubman, Goetz, Vincent, Yang, Yang, Zhang, and Zhao]{2018KIM_qmcpack}
Kim,~J.; Baczewski,~A.~D.; Beaudet,~T.~D.; Benali,~A.; Bennett,~M.~C.; Berrill,~M.~A.; Blunt,~N.~S.; Borda,~E. J.~L.; Casula,~M.; Ceperley,~D.~M.; Chiesa,~S.; Clark,~B.~K.; Clay,~R.~C.; Delaney,~K.~T.; Dewing,~M.; Esler,~K.~P.; Hao,~H.; Heinonen,~O.; Kent,~P. R.~C.; Krogel,~J.~T.; Kyl{\"a}np{\"a}{\"a},~I.; Li,~Y.~W.; Lopez,~M.~G.; Luo,~Y.; Malone,~F.~D.; Martin,~R.~M.; Mathuriya,~A.; McMinis,~J.; Melton,~C.~A.; Mitas,~L.; Morales,~M.~A.; Neuscamman,~E.; Parker,~W.~D.; Flores,~S. D.~P.; Romero,~N.~A.; Rubenstein,~B.~M.; Shea,~J. A.~R.; Shin,~H.; Shulenburger,~L.; Tillack,~A.~F.; Townsend,~J.~P.; Tubman,~N.~M.; Goetz,~B. V.~D.; Vincent,~J.~E.; Yang,~D.~C.; Yang,~Y.; Zhang,~S.; Zhao,~L. {QMCPACK}: an open sourceab initioquantum Monte Carlo package for the electronic structure of atoms, molecules and solids. \emph{J. Phys. Condens. Matter} \textbf{2018}, \emph{30}, 195901\relax
\mciteBstWouldAddEndPuncttrue
\mciteSetBstMidEndSepPunct{\mcitedefaultmidpunct}
{\mcitedefaultendpunct}{\mcitedefaultseppunct}\relax
\EndOfBibitem
\bibitem[Kent \latin{et~al.}(2020)Kent, Annaberdiyev, Benali, Bennett, Borda, Doak, Hao, Jordan, Krogel, Kylänpaä, Lee, Luo, Malone, Melton, Mitas, Morales, Neuscamman, Reboredo, Rubenstein, Saritas, Upadhyay, Wang, Zhang, and Zhao]{2020KEN_qmcpack}
Kent,~P.~R.; Annaberdiyev,~A.; Benali,~A.; Bennett,~M.~C.; Borda,~E. J.~L.; Doak,~P.; Hao,~H.; Jordan,~K.~D.; Krogel,~J.~T.; Kylänpaä,~I.; Lee,~J.; Luo,~Y.; Malone,~F.~D.; Melton,~C.~A.; Mitas,~L.; Morales,~M.~A.; Neuscamman,~E.; Reboredo,~F.~A.; Rubenstein,~B.; Saritas,~K.; Upadhyay,~S.; Wang,~G.; Zhang,~S.; Zhao,~L. QMCPACK: Advances in the development, efficiency, and application of auxiliary field and real-space variational and diffusion quantum Monte Carlo. \emph{J. Chem. Phys.} \textbf{2020}, \emph{152}\relax
\mciteBstWouldAddEndPuncttrue
\mciteSetBstMidEndSepPunct{\mcitedefaultmidpunct}
{\mcitedefaultendpunct}{\mcitedefaultseppunct}\relax
\EndOfBibitem
\bibitem[Neese and Valeev(2011)Neese, and Valeev]{neeseRevisitingAtomicNatural2011}
Neese,~F.; Valeev,~E.~F. Revisiting the {{Atomic Natural Orbital Approach}} for {{Basis Sets}}: {{Robust Systematic Basis Sets}} for {{Explicitly Correlated}} and {{Conventional Correlated}} {\emph{Ab Initio}} {{Methods}}? \emph{J. Chem. Theory Comput.} \textbf{2011}, \emph{7}, 33--43\relax
\mciteBstWouldAddEndPuncttrue
\mciteSetBstMidEndSepPunct{\mcitedefaultmidpunct}
{\mcitedefaultendpunct}{\mcitedefaultseppunct}\relax
\EndOfBibitem
\bibitem[Nakano \latin{et~al.}(2023)Nakano, Kohulák, Raghav, Casula, and Sorella]{2023NAK}
Nakano,~K.; Kohulák,~O.; Raghav,~A.; Casula,~M.; Sorella,~S. {TurboGenius: Python suite for high-throughput calculations of ab initio quantum Monte Carlo methods}. \emph{J. Chem. Phys.} \textbf{2023}, \emph{159}, 224801\relax
\mciteBstWouldAddEndPuncttrue
\mciteSetBstMidEndSepPunct{\mcitedefaultmidpunct}
{\mcitedefaultendpunct}{\mcitedefaultseppunct}\relax
\EndOfBibitem
\bibitem[Burkatzki \latin{et~al.}(2007)Burkatzki, Filippi, and Dolg]{2007burkatzkiecp}
Burkatzki,~M.; Filippi,~C.; Dolg,~M. Energy-consistent pseudopotentials for quantum Monte Carlo calculations. \emph{J. Chem. Phys.} \textbf{2007}, \emph{126}\relax
\mciteBstWouldAddEndPuncttrue
\mciteSetBstMidEndSepPunct{\mcitedefaultmidpunct}
{\mcitedefaultendpunct}{\mcitedefaultseppunct}\relax
\EndOfBibitem
\bibitem[{\v{R}}ez{\'a}{\v{c}} \latin{et~al.}(2008){\v{R}}ez{\'a}{\v{c}}, Jure{\v{c}}ka, Riley, {\v{C}}ern{\`y}, Valdes, Pluh{\'a}{\v{c}}kov{\'a}, Berka, {\v{R}}ez{\'a}{\v{c}}, Pito{\v{n}}{\'a}k, Vondr{\'a}{\v{s}}ek, and Hobza]{2008REZ_BEGDB}
{\v{R}}ez{\'a}{\v{c}},~J.; Jure{\v{c}}ka,~P.; Riley,~K.~E.; {\v{C}}ern{\`y},~J.; Valdes,~H.; Pluh{\'a}{\v{c}}kov{\'a},~K.; Berka,~K.; {\v{R}}ez{\'a}{\v{c}},~T.; Pito{\v{n}}{\'a}k,~M.; Vondr{\'a}{\v{s}}ek,~J.; Hobza,~P. Quantum chemical benchmark energy and geometry database for molecular clusters and complex molecular systems (www. begdb. com): a users manual and examples. \emph{Collect. Czechoslov. Chem. Commun.} \textbf{2008}, \emph{73}, 1261--1270\relax
\mciteBstWouldAddEndPuncttrue
\mciteSetBstMidEndSepPunct{\mcitedefaultmidpunct}
{\mcitedefaultendpunct}{\mcitedefaultseppunct}\relax
\EndOfBibitem
\end{mcitethebibliography}


\providecommand{\latin}[1]{#1}
\makeatletter
\providecommand{\doi}
  {\begingroup\let\do\@makeother\dospecials
  \catcode`\{=1 \catcode`\}=2 \doi@aux}
\providecommand{\doi@aux}[1]{\endgroup\texttt{#1}}
\makeatother
\providecommand*\mcitethebibliography{\thebibliography}
\csname @ifundefined\endcsname{endmcitethebibliography}  {\let\endmcitethebibliography\endthebibliography}{}
\begin{mcitethebibliography}{12}
\providecommand*\natexlab[1]{#1}
\providecommand*\mciteSetBstSublistMode[1]{}
\providecommand*\mciteSetBstMaxWidthForm[2]{}
\providecommand*\mciteBstWouldAddEndPuncttrue
  {\def\EndOfBibitem{\unskip.}}
\providecommand*\mciteBstWouldAddEndPunctfalse
  {\let\EndOfBibitem\relax}
\providecommand*\mciteSetBstMidEndSepPunct[3]{}
\providecommand*\mciteSetBstSublistLabelBeginEnd[3]{}
\providecommand*\EndOfBibitem{}
\mciteSetBstSublistMode{f}
\mciteSetBstMaxWidthForm{subitem}{(\alph{mcitesubitemcount})}
\mciteSetBstSublistLabelBeginEnd
  {\mcitemaxwidthsubitemform\space}
  {\relax}
  {\relax}

\bibitem[Kim \latin{et~al.}(2018)Kim, Baczewski, Beaudet, Benali, Bennett, Berrill, Blunt, Borda, Casula, Ceperley, Chiesa, Clark, Clay, Delaney, Dewing, Esler, Hao, Heinonen, Kent, Krogel, Kyl{\"a}np{\"a}{\"a}, Li, Lopez, Luo, Malone, Martin, Mathuriya, McMinis, Melton, Mitas, Morales, Neuscamman, Parker, Flores, Romero, Rubenstein, Shea, Shin, Shulenburger, Tillack, Townsend, Tubman, Goetz, Vincent, Yang, Yang, Zhang, and Zhao]{2018KIM_qmcpack}
Kim,~J.; Baczewski,~A.~D.; Beaudet,~T.~D.; Benali,~A.; Bennett,~M.~C.; Berrill,~M.~A.; Blunt,~N.~S.; Borda,~E. J.~L.; Casula,~M.; Ceperley,~D.~M.; Chiesa,~S.; Clark,~B.~K.; Clay,~R.~C.; Delaney,~K.~T.; Dewing,~M.; Esler,~K.~P.; Hao,~H.; Heinonen,~O.; Kent,~P. R.~C.; Krogel,~J.~T.; Kyl{\"a}np{\"a}{\"a},~I.; Li,~Y.~W.; Lopez,~M.~G.; Luo,~Y.; Malone,~F.~D.; Martin,~R.~M.; Mathuriya,~A.; McMinis,~J.; Melton,~C.~A.; Mitas,~L.; Morales,~M.~A.; Neuscamman,~E.; Parker,~W.~D.; Flores,~S. D.~P.; Romero,~N.~A.; Rubenstein,~B.~M.; Shea,~J. A.~R.; Shin,~H.; Shulenburger,~L.; Tillack,~A.~F.; Townsend,~J.~P.; Tubman,~N.~M.; Goetz,~B. V.~D.; Vincent,~J.~E.; Yang,~D.~C.; Yang,~Y.; Zhang,~S.; Zhao,~L. {QMCPACK}: an open sourceab initioquantum Monte Carlo package for the electronic structure of atoms, molecules and solids. \emph{J. Phys. Condens. Matter} \textbf{2018}, \emph{30}, 195901\relax
\mciteBstWouldAddEndPuncttrue
\mciteSetBstMidEndSepPunct{\mcitedefaultmidpunct}
{\mcitedefaultendpunct}{\mcitedefaultseppunct}\relax
\EndOfBibitem
\bibitem[Kent \latin{et~al.}(2020)Kent, Annaberdiyev, Benali, Bennett, Borda, Doak, Hao, Jordan, Krogel, Kylänpaä, Lee, Luo, Malone, Melton, Mitas, Morales, Neuscamman, Reboredo, Rubenstein, Saritas, Upadhyay, Wang, Zhang, and Zhao]{2020KEN_qmcpack}
Kent,~P.~R.; Annaberdiyev,~A.; Benali,~A.; Bennett,~M.~C.; Borda,~E. J.~L.; Doak,~P.; Hao,~H.; Jordan,~K.~D.; Krogel,~J.~T.; Kylänpaä,~I.; Lee,~J.; Luo,~Y.; Malone,~F.~D.; Melton,~C.~A.; Mitas,~L.; Morales,~M.~A.; Neuscamman,~E.; Reboredo,~F.~A.; Rubenstein,~B.; Saritas,~K.; Upadhyay,~S.; Wang,~G.; Zhang,~S.; Zhao,~L. QMCPACK: Advances in the development, efficiency, and application of auxiliary field and real-space variational and diffusion quantum Monte Carlo. \emph{J. Chem. Phys.} \textbf{2020}, \emph{152}\relax
\mciteBstWouldAddEndPuncttrue
\mciteSetBstMidEndSepPunct{\mcitedefaultmidpunct}
{\mcitedefaultendpunct}{\mcitedefaultseppunct}\relax
\EndOfBibitem
\bibitem[Bennett \latin{et~al.}(2017)Bennett, Melton, Annaberdiyev, Wang, Shulenburger, and Mitas]{2017BEN}
Bennett,~M.~C.; Melton,~C.~A.; Annaberdiyev,~A.; Wang,~G.; Shulenburger,~L.; Mitas,~L. {A new generation of effective core potentials for correlated calculations}. \emph{J. Chem. Phys} \textbf{2017}, \emph{147}, 224106\relax
\mciteBstWouldAddEndPuncttrue
\mciteSetBstMidEndSepPunct{\mcitedefaultmidpunct}
{\mcitedefaultendpunct}{\mcitedefaultseppunct}\relax
\EndOfBibitem
\bibitem[Bennett \latin{et~al.}(2018)Bennett, Wang, Annaberdiyev, Melton, Shulenburger, and Mitas]{2018BEN}
Bennett,~M.~C.; Wang,~G.; Annaberdiyev,~A.; Melton,~C.~A.; Shulenburger,~L.; Mitas,~L. {A new generation of effective core potentials from correlated calculations: 2nd row elements}. \emph{J. Chem. Phys.} \textbf{2018}, \emph{149}, 104108\relax
\mciteBstWouldAddEndPuncttrue
\mciteSetBstMidEndSepPunct{\mcitedefaultmidpunct}
{\mcitedefaultendpunct}{\mcitedefaultseppunct}\relax
\EndOfBibitem
\bibitem[Perdew and Zunger(1981)Perdew, and Zunger]{1981PER}
Perdew,~J.~P.; Zunger,~A. Self-interaction correction to density-functional approximations for many-electron systems. \emph{Phys. Rev. B} \textbf{1981}, \emph{23}, 5048\relax
\mciteBstWouldAddEndPuncttrue
\mciteSetBstMidEndSepPunct{\mcitedefaultmidpunct}
{\mcitedefaultendpunct}{\mcitedefaultseppunct}\relax
\EndOfBibitem
\bibitem[Sun \latin{et~al.}(2018)Sun, Berkelbach, Blunt, Booth, Guo, Li, Liu, McClain, Sayfutyarova, Sharma, \latin{et~al.} others]{2018SUN}
Sun,~Q.; Berkelbach,~T.~C.; Blunt,~N.~S.; Booth,~G.~H.; Guo,~S.; Li,~Z.; Liu,~J.; McClain,~J.~D.; Sayfutyarova,~E.~R.; Sharma,~S.; others PySCF: the Python-based simulations of chemistry framework. \emph{Wiley Interdiscip. Rev. Comput. Mol. Sci.} \textbf{2018}, \emph{8}, e1340\relax
\mciteBstWouldAddEndPuncttrue
\mciteSetBstMidEndSepPunct{\mcitedefaultmidpunct}
{\mcitedefaultendpunct}{\mcitedefaultseppunct}\relax
\EndOfBibitem
\bibitem[Sun \latin{et~al.}(2020)Sun, Zhang, Banerjee, Bao, Barbry, Blunt, Bogdanov, Booth, Chen, Cui, Eriksen, Gao, Guo, Hermann, Hermes, Koh, Koval, Lehtola, Li, Liu, Mardirossian, McClain, Motta, Mussard, Pham, Pulkin, Purwanto, Robinson, Ronca, Sayfutyarova, Scheurer, Schurkus, Smith, Sun, Sun, Upadhyay, Wagner, Wang, White, Whitfield, Williamson, Wouters, Yang, Yu, Zhu, Berkelbach, Sharma, Sokolov, and Chan]{2020SUN}
Sun,~Q.; Zhang,~X.; Banerjee,~S.; Bao,~P.; Barbry,~M.; Blunt,~N.~S.; Bogdanov,~N.~A.; Booth,~G.~H.; Chen,~J.; Cui,~Z.~H.; Eriksen,~J.~J.; Gao,~Y.; Guo,~S.; Hermann,~J.; Hermes,~M.~R.; Koh,~K.; Koval,~P.; Lehtola,~S.; Li,~Z.; Liu,~J.; Mardirossian,~N.; McClain,~J.~D.; Motta,~M.; Mussard,~B.; Pham,~H.~Q.; Pulkin,~A.; Purwanto,~W.; Robinson,~P.~J.; Ronca,~E.; Sayfutyarova,~E.~R.; Scheurer,~M.; Schurkus,~H.~F.; Smith,~J.~E.; Sun,~C.; Sun,~S.~N.; Upadhyay,~S.; Wagner,~L.~K.; Wang,~X.; White,~A.; Whitfield,~J.~D.; Williamson,~M.~J.; Wouters,~S.; Yang,~J.; Yu,~J.~M.; Zhu,~T.; Berkelbach,~T.~C.; Sharma,~S.; Sokolov,~A.~Y.; Chan,~G. K.~L. Recent developments in the PySCF program package. \emph{J. Chem. Phys.} \textbf{2020}, \emph{153}, 024109\relax
\mciteBstWouldAddEndPuncttrue
\mciteSetBstMidEndSepPunct{\mcitedefaultmidpunct}
{\mcitedefaultendpunct}{\mcitedefaultseppunct}\relax
\EndOfBibitem
\bibitem[Giannozzi \latin{et~al.}(2009)Giannozzi, Baroni, Bonini, Calandra, Car, Cavazzoni, Ceresoli, Chiarotti, Cococcioni, Dabo, \latin{et~al.} others]{2009GIA_QE}
Giannozzi,~P.; Baroni,~S.; Bonini,~N.; Calandra,~M.; Car,~R.; Cavazzoni,~C.; Ceresoli,~D.; Chiarotti,~G.~L.; Cococcioni,~M.; Dabo,~I.; others QUANTUM ESPRESSO: a modular and open-source software project for quantum simulations of materials. \emph{J. Phys. Condens. Matter.} \textbf{2009}, \emph{21}, 395502\relax
\mciteBstWouldAddEndPuncttrue
\mciteSetBstMidEndSepPunct{\mcitedefaultmidpunct}
{\mcitedefaultendpunct}{\mcitedefaultseppunct}\relax
\EndOfBibitem
\bibitem[Alf\`e and Gillan(2004)Alf\`e, and Gillan]{splines}
Alf\`e,~D.; Gillan,~M.~J. Efficient localized basis set for quantum Monte Carlo calculations on condensed matter. \emph{Phys. Rev. B} \textbf{2004}, \emph{70}, 161101\relax
\mciteBstWouldAddEndPuncttrue
\mciteSetBstMidEndSepPunct{\mcitedefaultmidpunct}
{\mcitedefaultendpunct}{\mcitedefaultseppunct}\relax
\EndOfBibitem
\bibitem[Casula(2006)]{2006CAS_tmove}
Casula,~M. Beyond the Locality Approximation in the Standard Diffusion {{Monte Carlo}} Method. \emph{Phys. Rev. B} \textbf{2006}, \emph{74}, 161102\relax
\mciteBstWouldAddEndPuncttrue
\mciteSetBstMidEndSepPunct{\mcitedefaultmidpunct}
{\mcitedefaultendpunct}{\mcitedefaultseppunct}\relax
\EndOfBibitem
\bibitem[Zen \latin{et~al.}(2016)Zen, Sorella, Gillan, Michaelides, and Alf{\`{e}}]{Zen2016}
Zen,~A.; Sorella,~S.; Gillan,~M.~J.; Michaelides,~A.; Alf{\`{e}},~D. {Boosting the accuracy and speed of quantum Monte Carlo: Size consistency and time step}. \emph{Phys. Rev. B} \textbf{2016}, \emph{93}, 241118(R)\relax
\mciteBstWouldAddEndPuncttrue
\mciteSetBstMidEndSepPunct{\mcitedefaultmidpunct}
{\mcitedefaultendpunct}{\mcitedefaultseppunct}\relax
\EndOfBibitem
\end{mcitethebibliography}

\end{document}


\begin{abstract}
This file includes the supplementary information for the paper titled ``Basis set incompleteness errors in fixed-node diffusion Monte Carlo calculations on non-covalent interactions".
\end{abstract}

\makeatletter
\def\Hline{
\noalign{\ifnum0=`}\fi\hrule \@height 1pt \futurelet
\reserved@a\@xhline}
\makeatother

\makeatletter
\renewcommand{\refname}{}
\renewcommand*{\citenumfont}[1]{#1}
\renewcommand*{\bibnumfmt}[1]{[#1]}
\makeatother

\setcounter{table}{0}
\setcounter{equation}{0}
\setcounter{figure}{0}
\renewcommand{\thepage}{s\arabic{page}}
\renewcommand{\thetable}{S\Roman{table}}
\renewcommand{\thefigure}{S\arabic{figure}}
\renewcommand{\theequation}{S\arabic{equation}}

\listoffigures
\listoftables

\newpage
\section{Additional QMCPACK calculations}
\label{sec:qmcpack}

QMCPACK~\cite{2018KIM_qmcpack, 2020KEN_qmcpack} is a high-performance real space QMC code capable of performing molecular and solid state calculations on modern CPU and GPU machines.
QMCPACK implements the standard (time-discretized) FN-DMC algorithm, as well as the variational QMC algorithm and several schemes for the trial wavefunction optimization.

QMCPACK simulations were performed using a Slater-Jastrow trial wavefunction, with the determinant part of the wavefunction being a single Slater determinant, and employing the ccECP pseudopotentials~\cite{2017BEN,2018BEN} (as for to the TurboRVB simulations). 
The Slater determinant is obtained from a preliminary DFT calculation employing the PZ-LDA~{\cite{1981PER}} exchange-correlation functional.
We performed two sets of simulations, using localized GTO basis sets or PW basis sets to express the DFT single-particle orbitals. 
Five sets of calculations were done for the GTO trial wavefunctions, using the following basis sets (designed for the ccECP pseudopotentials): ccecp-cc-pVDZ (VDZ), ccecp-aug-cc-pVDZ (aVDZ), ccecp-cc-pVTZ (VTZ),  ccecp-aug-cc-pVTZ (aVTZ), and ccecp-aug-cc-pV6Z (aV6Z). The preliminary DFT calculation was performed  using the \pyscf~{\cite{2018SUN, 2020SUN}} package.
To generate plane-wave (PW) based trial wavefunctions 
we used the \qe~\cite{2009GIA_QE} package with a $600\,$Ryd energy cutoff. The single-particle orbitals obtained are then reexpanded in terms of B-splines~\cite{splines} in the QMC calculations.

We used the default Jastrow factor implemented in QMCPACK, which includes electron-nucleus, electron-electron, and electron-electron-nucleus terms.
The electron-nucleus and electron-electron functions are both one-dimensional B-spline (tricubic spline on a linear grid) between zero and a cutoff distance.
The electron-electron-nucleus function is a polynomial expansion.
The parameters of the Jastrow factor have been optimised by minimising the variational energy of each system, employing the linear method with line minimization via quartic polynomial fits, and performing several steps of optimization with a sampling of up to 10,000,000 configurations.

The DMC calculations were performed with the T-move localization scheme\cite{2006CAS_tmove}. 
DMC simulations have been performed with a target population of 102,400 walkers, and employing the modification to the drift and branching terms suggested in Ref.~\citenum{Zen2016}, called with the flag  \verb|ZSGMA|.
The reason for using the T-move localization scheme in QMCPACK calculations, instead of the DTM scheme used in the TurboRVB evaluations is that the available version (version 3.17.1, released on 25 August 2023) and previous versions of QMCPACK are affected by a bug on the DTM implementation.
Note that the different scheme used for the non-local psudopotential terms in QMCPACK and TurboRVB calculations implies that the QMCPACK-based evaluations (i.e., with T-move) and TurboRVB-based evaluations (i.e., with DTM) are not equivalent. 
The reported DMC results are obtained for a timestep $\tau=0.01\,$au.
Some additional DMC simulations with $\tau=0.03\,$au and $\tau=0.003\,$au show that the $\tau=0.01\,$au evaluations of the binding energy have a negligible timestep bias (namely, the timestep bias is smaller than the stochastic error). 

\begin{figure*}[htbp]
\centering
\includegraphics[width=1.0\textwidth]{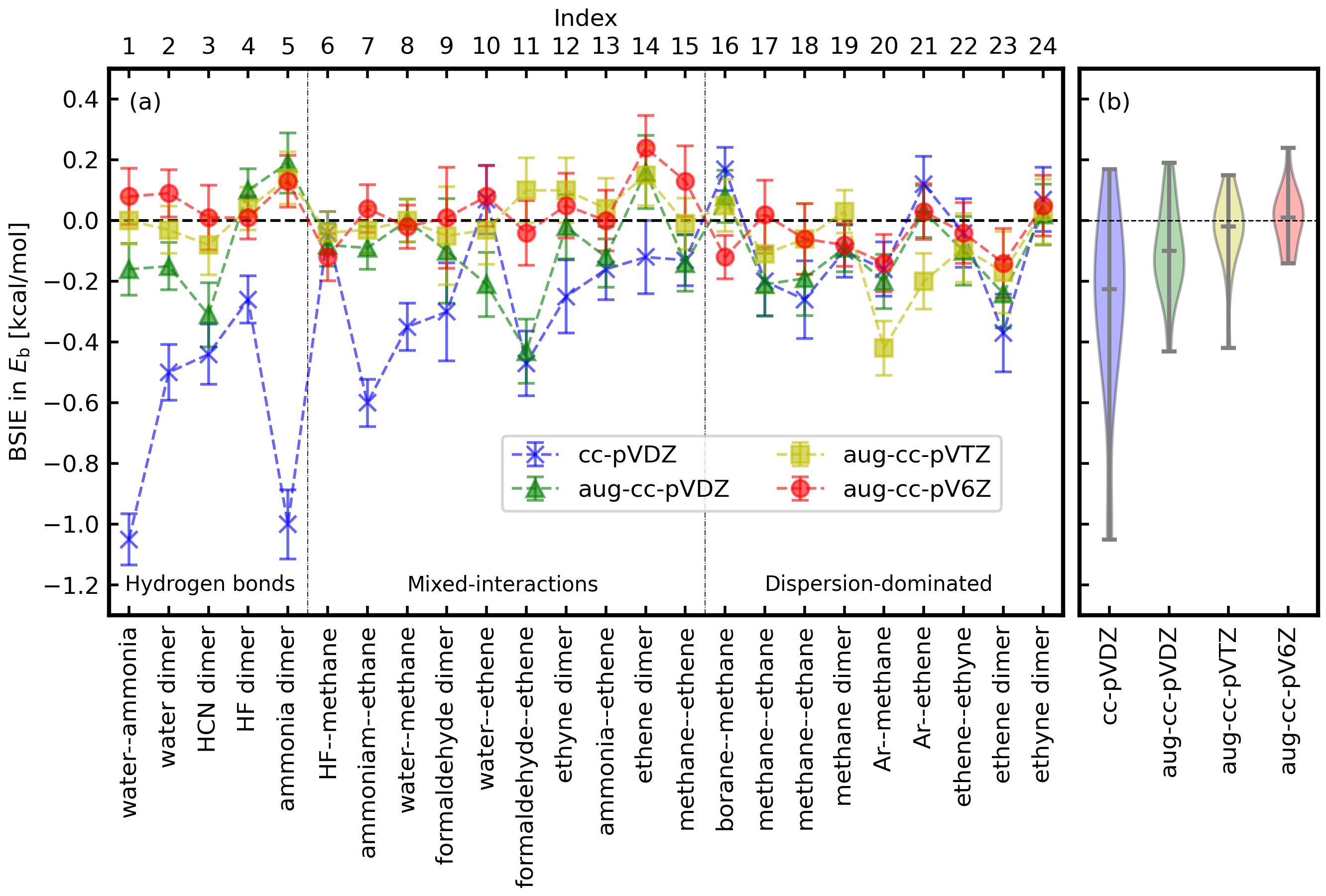}
\caption{(a) BSIEs in the binding energies of the A24 set, estimated from DMC calculations with QMCPACK, where the reference values are those obtained with PW basis (i.e., the CBS limit). The error bars represent $1\sigma$. (b) The violin plots for the obtained BSIEs. In panel (b), the mean values of the obtained binding energies for computing the distribution densities.}
\label{fig:SI-BSIE-summary-qmcpack}
\end{figure*}

DMC evaluations of the  binding energies obtained using the above setup are reported in Table \ref{tab:qmcpack}.
The BSIE of the evaluations are shown in Figure~\ref{fig:SI-BSIE-summary-qmcpack}.

\begin{center}
\begin{table}[hbtp]
\caption{\label{tab:qmcpack}
The binding energies (in kcal/mol) of the molecule contained in the A24-set, as obtained from FN-DMC evaluations using QMCPACK and the setup described in section~\ref{sec:qmcpack}.
}
\begin{tabular}{lrrrrrr}
\hline
 &  VDZ &  VTZ & aVDZ & aVTZ & aV6Z &            PW \\
\hline
      water--ammonia &     -7.74(6) &     -7.15(6) &        -6.85(6) &        -6.69(5) &        -6.61(7) &  -6.69(6) \\
         water dimer &     -5.74(7) &     -5.37(6) &        -5.39(5) &        -5.27(5) &        -5.15(5) &  -5.24(6) \\
           HCN dimer &     -5.48(7) &    -5.39(10) &        -5.35(8) &        -5.12(7) &        -5.03(8) &  -5.04(7) \\
            HF dimer &     -5.08(6) &     -4.86(6) &        -4.72(5) &        -4.78(5) &        -4.81(5) &  -4.82(5) \\
       ammonia dimer &     -4.30(9) &     -3.56(6) &        -3.11(7) &        -3.16(5) &        -3.17(5) &  -3.30(7) \\
         HF--methane &     -1.62(5) &     -1.82(9) &        -1.66(5) &        -1.62(5) &        -1.70(6) &  -1.58(5) \\
    ammoniam--ethane &     -1.40(6) &     -1.11(5) &        -0.89(5) &        -0.83(6) &        -0.76(6) &  -0.80(5) \\
      water--methane &     -1.02(6) &     -0.75(6) &        -0.67(5) &        -0.67(5) &        -0.69(5) &  -0.67(5) \\
  formaldehyde dimer &     -4.89(8) &     -4.80(8) &       -4.69(10) &        -4.64(8) &        -4.58(9) & -4.59(14) \\
       water--ethene &     -2.55(8) &     -2.80(8) &        -2.83(7) &        -2.65(8) &        -2.54(6) &  -2.62(8) \\
formaldehyde--ethene &     -2.03(8) &     -1.88(8) &        -1.99(8) &        -1.46(8) &        -1.60(8) &  -1.56(7) \\
        ethyne dimer &     -1.86(9) &     -1.54(7) &        -1.63(7) &        -1.51(7) &        -1.56(7) &  -1.61(8) \\
     ammonia--ethene &     -1.62(8) &     -1.59(7) &        -1.58(8) &        -1.42(8) &        -1.46(8) &  -1.46(6) \\
        ethene dimer &     -1.25(9) &     -1.07(8) &        -0.97(9) &        -0.98(6) &        -0.89(7) &  -1.13(8) \\
     methane--ethene &     -0.70(6) &     -0.48(7) &        -0.71(7) &        -0.58(6) &       -0.44(10) &  -0.57(6) \\
     borane--methane &     -1.22(5) &     -1.41(5) &        -1.31(7) &        -1.34(7) &        -1.51(5) &  -1.39(5) \\
     methane--ethane &     -0.92(7) &     -0.95(8) &        -0.93(5) &        -0.83(9) &        -0.70(7) &  -0.72(9) \\
     methane--ethane &     -0.83(8) &     -0.57(9) &        -0.76(7) &        -0.63(6) &        -0.63(6) & -0.57(10) \\
       methane dimer &     -0.64(7) &     -0.55(4) &        -0.63(6) &        -0.51(5) &        -0.62(5) &  -0.54(5) \\
         Ar--methane &     -0.17(4) &     -0.07(4) &        -0.21(4) &        -0.43(4) &        -0.15(5) &  -0.01(8) \\
          Ar--ethene &      0.36(7) &      0.18(8) &         0.27(6) &         0.04(7) &         0.27(7) &   0.24(6) \\
      ethene--ethyne &      1.09(8) &      1.01(9) &         1.03(8) &         1.04(8) &         1.09(6) &   1.13(8) \\
        ethene dimer &      1.00(9) &      1.17(8) &         1.13(7) &        1.20(10) &         1.23(7) &   1.37(9) \\
        ethyne dimer &      1.27(7) &      1.28(9) &         1.22(6) &         1.23(7) &         1.25(6) &   1.20(8) \\
\hline
\end{tabular}
\end{table}
\end{center}

\clearpage
\section{Supporting Results}

\begin{center}
\begin{table}[hbtp]
\caption{\label{tab:turbo-ccpvxz}
The binding energies (in kcal/mol) of the molecule contained in the A24-set, as obtained from FN-DMC evaluations using TurboRVB with the ccecp-cc-pV$n$Z (V$n$Z) basis set family ($n$=D, T, Q, 5, 6) using the setup described in the main text.
}
\begin{tabular}{lrrrrr}
\hline
Label & VDZ & VTZ & VQZ & V5Z & V6Z \\
\hline
water--ammonia & -7.63(7) & -7.13(6) & -6.83(7) & -6.76(7) & -6.68(7) \\
water dimer & -5.73(7) & -5.47(7) & -5.29(7) & -5.15(7) & -5.19(7) \\
HCN dimer & -5.56(7) & -5.16(7) & -5.16(7) & -5.09(8) & -5.22(7) \\
HF dimer & -4.91(7) & -4.79(7) & -4.64(7) & -4.64(7) & -4.77(7) \\
ammonia dimer & -4.43(7) & -3.60(7) & -3.18(7) & -3.22(7) & -3.12(7) \\
HF--methane & -1.62(7) & -1.59(6) & -1.59(7) & -1.63(7) & -1.59(6) \\
ammoniam--ethane & -1.31(7) & -1.03(7) & -0.81(7) & -0.76(6) & -0.76(7) \\
water--methane & -0.94(8) & -0.57(7) & -0.86(6) & -0.62(6) & -0.65(7) \\
formaldehyde dimer & -4.78(6) & -4.79(8) & -4.63(7) & -4.56(9) & -4.60(9) \\
water--ethene & -2.69(8) & -2.79(7) & -2.64(6) & -2.71(6) & -2.65(7) \\
formaldehyde--ethene & -1.89(8) & -1.88(7) & -1.64(7) & -1.63(8) & -1.58(8) \\
ethyne dimer & -1.78(7) & -1.52(6) & -1.58(6) & -1.59(7) & -1.56(7) \\
ammonia--ethene & -1.43(6) & -1.39(7) & -1.46(7) & -1.42(8) & -1.38(7) \\
ethene dimer & -1.13(7) & -1.05(7) & -0.99(7) & -0.89(8) & -1.08(7) \\
methane--ethene & -0.55(8) & -0.47(6) & -0.47(8) & -0.55(6) & -0.48(7) \\
borane--methane & -1.16(7) & -1.35(7) & -1.49(6) & -1.23(7) & -1.33(6) \\
methane--ethane & -0.77(8) & -0.79(7) & -0.72(7) & -0.77(7) & -0.88(7) \\
methane--ethane & -0.64(7) & -0.43(7) & -0.64(6) & -0.68(6) & -0.51(6) \\
methane dimer & -0.46(7) & -0.55(7) & -0.48(6) & -0.38(7) & -0.47(6) \\
Ar--methane & -0.39(7) & -0.31(5) & -0.40(6) & -0.42(8) & -0.48(7) \\
Ar--ethene & -0.38(6) & -0.37(7) & -0.29(7) & -0.36(7) & -0.22(7) \\
ethene--ethyne & 1.10(7) & 1.08(6) & 0.99(7) & 0.91(7) & 0.99(6) \\
ethene dimer & 1.05(7) & 0.95(7) & 1.13(8) & 1.16(7) & 1.16(8) \\
ethyne dimer & 1.33(7) & 1.25(7) & 1.17(7) & 1.27(7) & 1.45(6) \\
\hline
\end{tabular}
\end{table}
\end{center}

\begin{center}
\begin{table}[hbtp]
\caption{\label{tab:turbo-ccpvxz}
The binding energies (in kcal/mol) of the molecule contained in the A24-set, as obtained from FN-DMC evaluations using TurboRVB with the ccecp-aug-cc-pV$n$Z (aV$n$Z) basis set family ($n$=D, T, Q, 5, 6) using the setup described in the main text.
}
\begin{tabular}{lrrrrr}
\hline
 & aVDZ & aVTZ & aVQZ & aV5Z & aV6Z \\
\hline
water--ammonia & -6.73(7) & -6.61(6) & -6.65(6) & -6.64(6) & -6.75(7) \\
water dimer & -5.20(6) & -5.12(7) & -5.08(7) & -5.19(6) & -5.10(8) \\
HCN dimer & -5.54(7) & -5.09(7) & -5.11(7) & -5.03(8) & -5.09(7) \\
HF dimer & -4.70(7) & -4.56(7) & -4.62(8) & -4.78(7) & -4.74(7) \\
ammonia dimer & -3.19(6) & -3.12(7) & -3.04(6) & -3.18(7) & -3.10(6) \\
HF--methane & -1.90(7) & -1.72(7) & -1.65(7) & -1.53(7) & -1.64(7) \\
ammoniam--ethane & -0.69(7) & -0.71(7) & -0.70(7) & -0.68(6) & -0.80(7) \\
water--methane & -0.77(7) & -0.67(6) & -0.65(7) & -0.66(7) & -0.58(6) \\
formaldehyde dimer & -4.57(7) & -4.71(7) & -4.72(9) & -4.64(8) & -4.42(9) \\
water--ethene & -2.64(7) & -2.68(8) & -2.53(7) & -2.68(8) & -2.50(10) \\
formaldehyde--ethene & -1.81(7) & -1.65(7) & -1.70(8) & -1.43(9) & -1.71(10) \\
ethyne dimer & -1.59(7) & -1.71(6) & -1.61(7) & -1.58(7) & -1.44(7) \\
ammonia--ethene & -1.51(6) & -1.57(6) & -1.33(6) & -1.27(7) & -1.38(6) \\
ethene dimer & -1.19(6) & -1.04(7) & -0.96(8) & -1.05(9) & -0.97(9) \\
methane--ethene & -0.53(7) & -0.51(7) & -0.36(7) & -0.49(7) & -0.56(6) \\
borane--methane & -1.44(6) & -1.36(7) & -1.45(7) & -1.30(5) & -1.46(7) \\
methane--ethane & -0.93(7) & -0.83(6) & -0.78(8) & -0.71(8) & -0.65(9) \\
methane--ethane & -0.86(7) & -0.48(8) & -0.49(7) & -0.58(7) & -0.57(8) \\
methane dimer & -0.59(7) & -0.54(7) & -0.51(7) & -0.51(7) & -0.58(6) \\
Ar--methane & -0.35(7) & -0.44(7) & -0.45(7) & -0.33(7) & -0.36(8) \\
Ar--ethene & -0.40(7) & -0.29(7) & -0.31(7) & -0.34(8) & -0.24(7) \\
ethene--ethyne & 1.03(8) & 1.02(7) & 0.98(8) & 0.84(8) & 1.04(9) \\
ethene dimer & 0.97(7) & 1.14(7) & 1.06(8) & 1.16(7) & 1.04(8) \\
ethyne dimer & 1.34(7) & 1.27(7) & 1.17(6) & 1.28(7) & 1.32(8) \\
\hline
\end{tabular}
\end{table}
\end{center}

\begin{figure*}[htbp]
\centering
\includegraphics[width=4.5in]{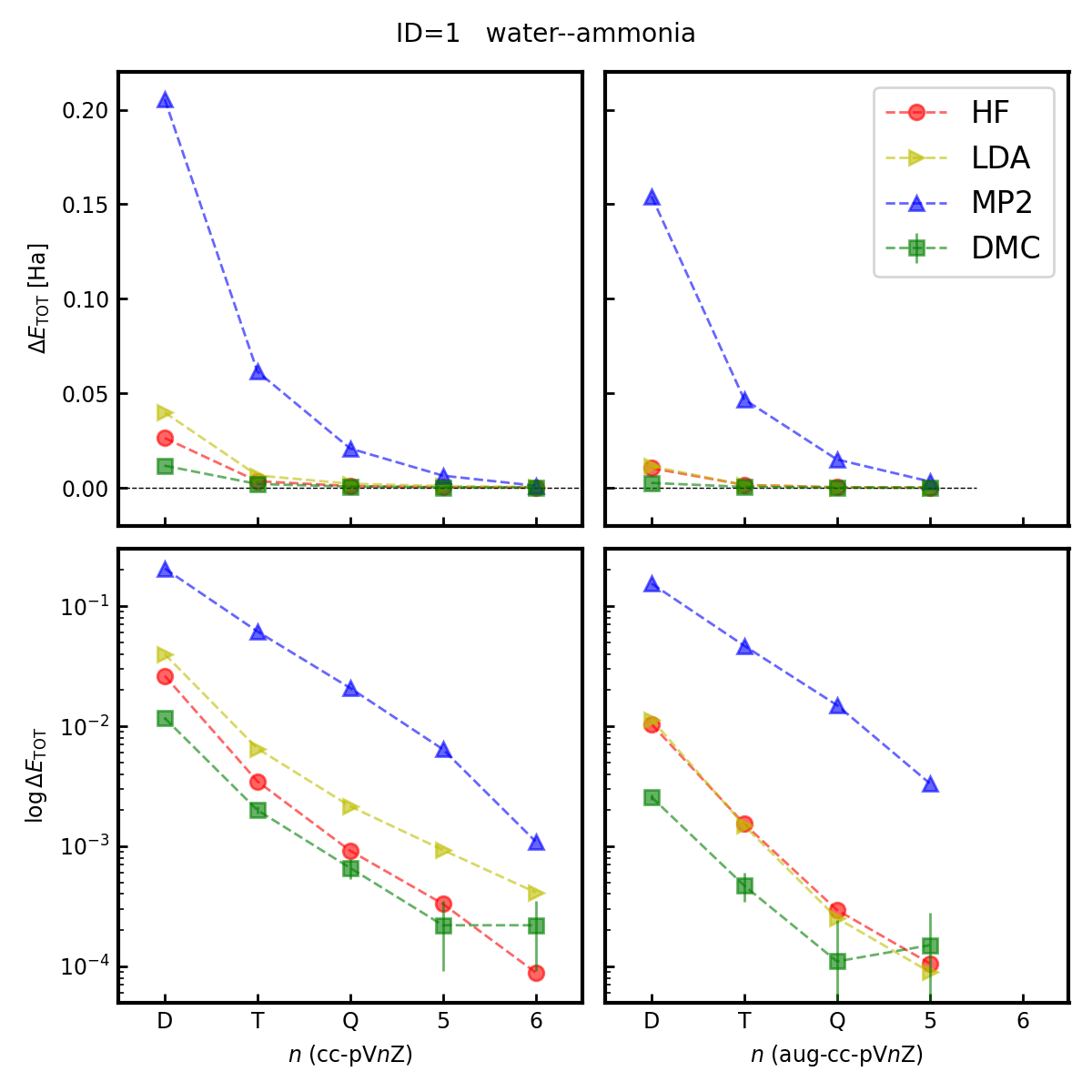}
\caption{Total Energy of the water--ammonia dimer in the A24 data set with respect to the energy obitained with the aug-cc-pV6Z basis set.}
\label{fig:SI-total-energy-01waterammonia}
\end{figure*}

\begin{figure*}[htbp]
\centering
\includegraphics[width=0.9\textwidth]{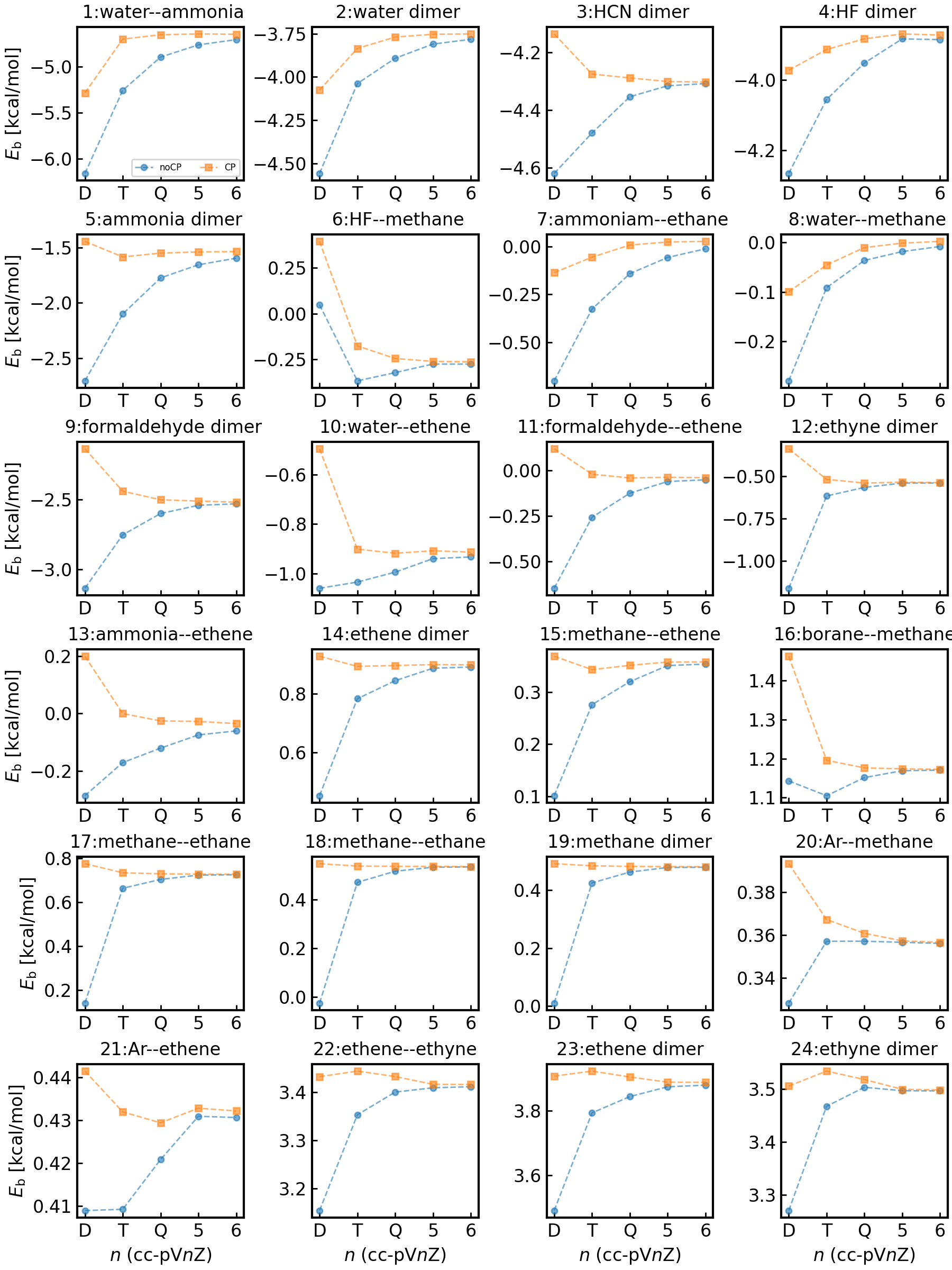}
\caption{The binding energies of A24 data set computed by HF with cc-pV$n$Z basis sets ($n$=D,T,Q,5,6). The CP and noCP represent `with the counterpoise correction' and `without the counterpoise correction', respectively.}
\label{fig:SI-binding-hf-ccpvnz}
\end{figure*}

\begin{figure*}[htbp]
\centering
\includegraphics[width=0.9\textwidth]{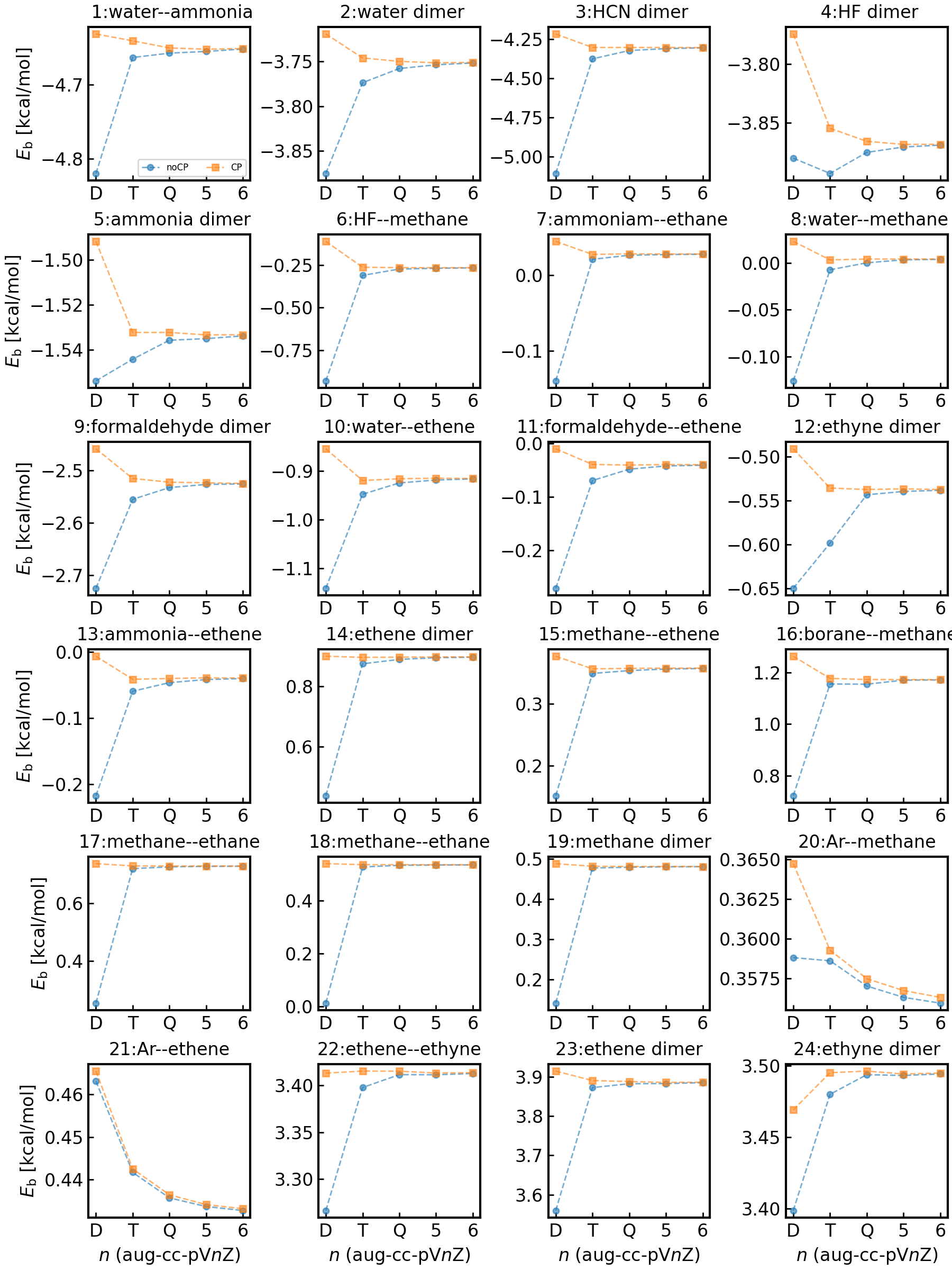}
\caption{The binding energies of A24 data set computed by HF with aug-cc-pV$n$Z basis sets ($n$=D,T,Q,5,6). The CP and noCP represent `with the counterpoise correction' and `without the counterpoise correction', respectively.}
\label{fig:SI-binding-hf-augccpvnz}
\end{figure*}

\begin{figure*}[htbp]
\centering
\includegraphics[width=0.9\textwidth]{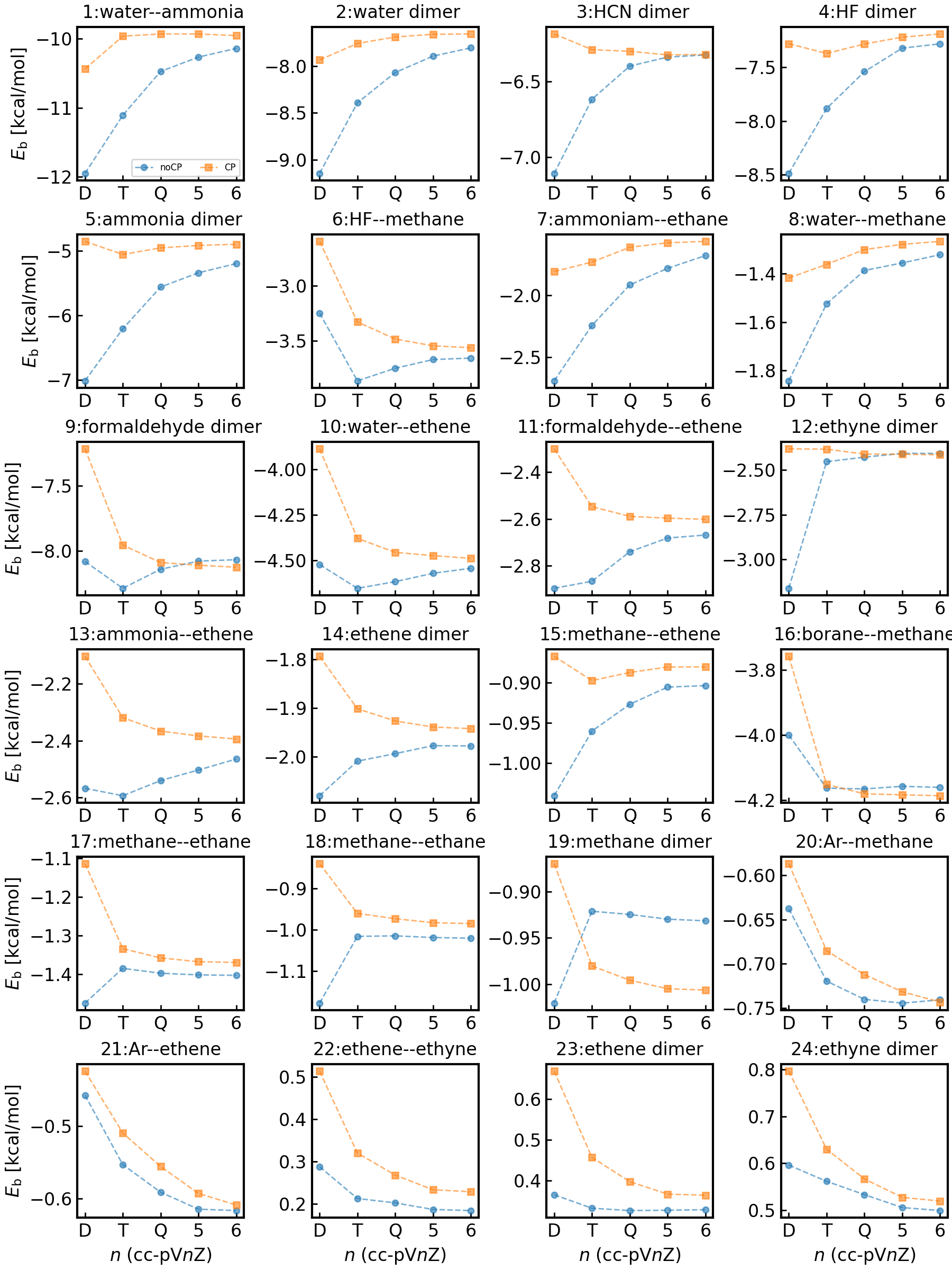}
\caption{The binding energies of A24 data set computed by LDA with cc-pV$n$Z basis sets ($n$=D,T,Q,5,6). The CP and noCP represent `with the counterpoise correction' and `without the counterpoise correction', respectively.}
\label{fig:SI-binding-lda-ccpvnz}
\end{figure*}

\begin{figure*}[htbp]
\centering
\includegraphics[width=0.9\textwidth]{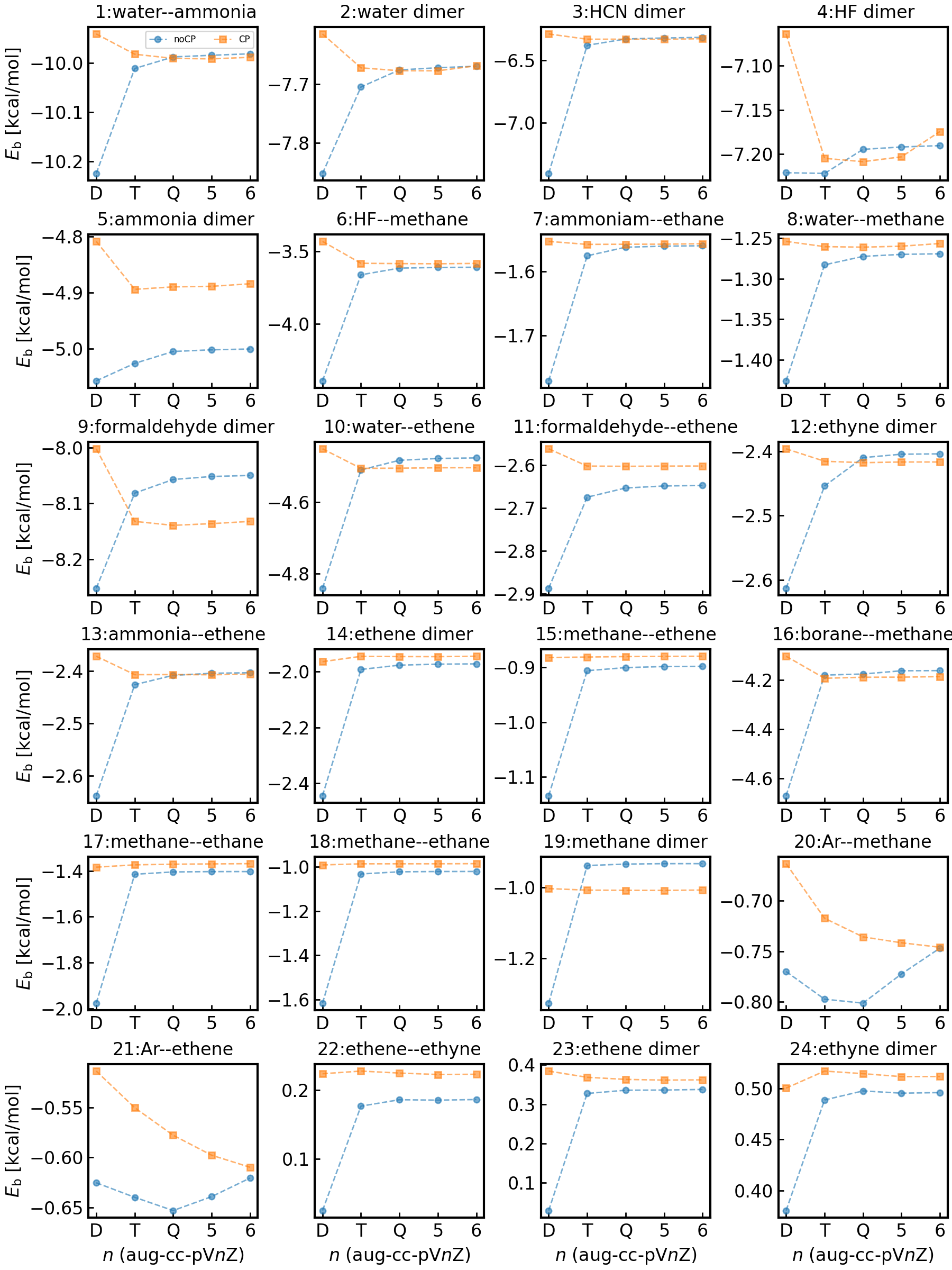}
\caption{The binding energies of A24 data set computed by LDA with aug-cc-pV$n$Z basis sets ($n$=D,T,Q,5,6). The CP and noCP represent `with the counterpoise correction' and `without the counterpoise correction', respectively.}
\label{fig:SI-binding-lda-augccpvnz}
\end{figure*}

\begin{figure*}[htbp]
\centering
\includegraphics[width=0.9\textwidth]{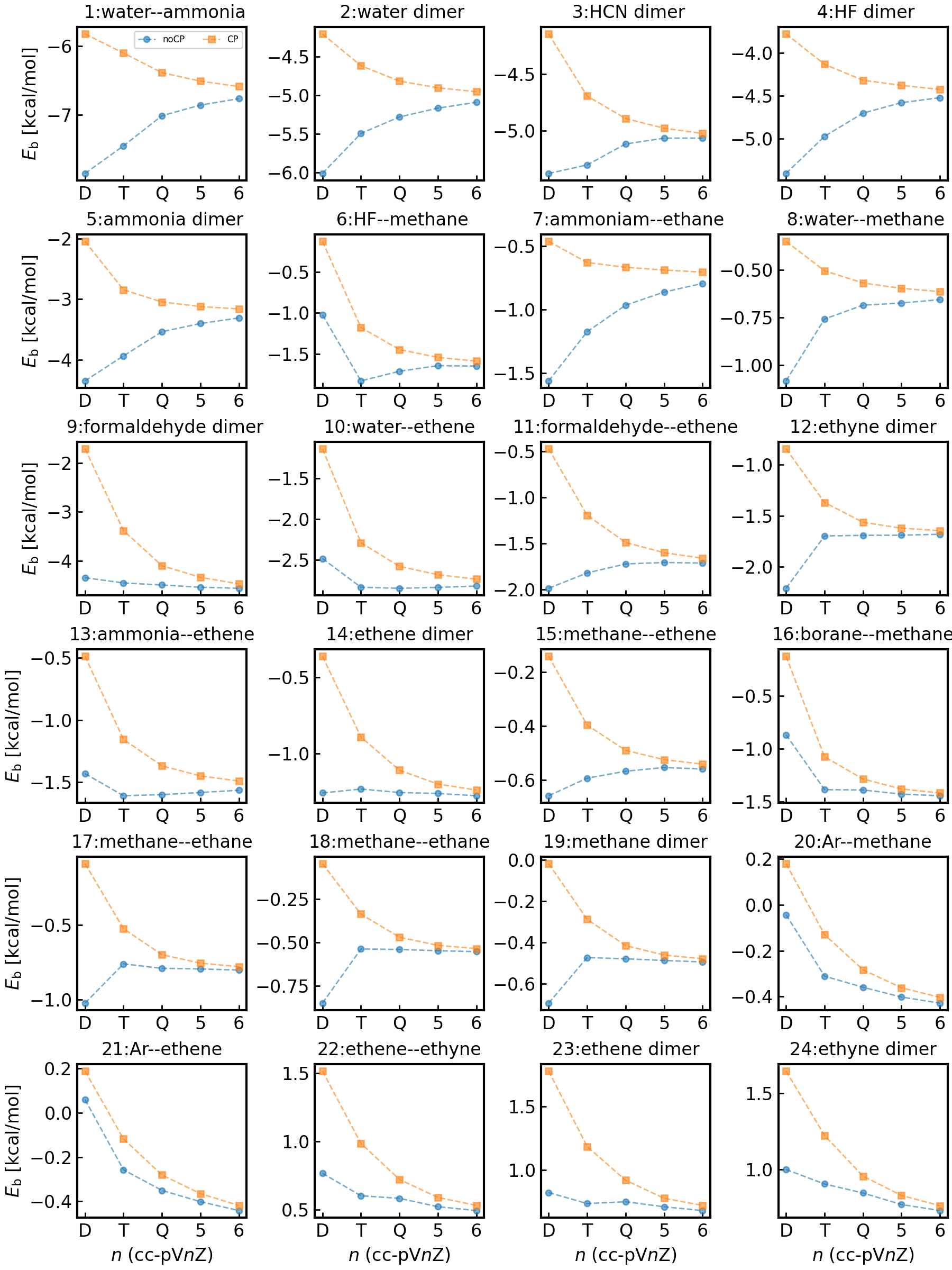}
\caption{The binding energies of A24 data set computed by MP2 with cc-pV$n$Z basis sets ($n$=D,T,Q,5,6). The CP and noCP represent `with the counterpoise correction' and `without the counterpoise correction', respectively.}
\label{fig:SI-binding-mp2-ccpvnz}
\end{figure*}

\begin{figure*}[htbp]
\centering
\includegraphics[width=0.9\textwidth]{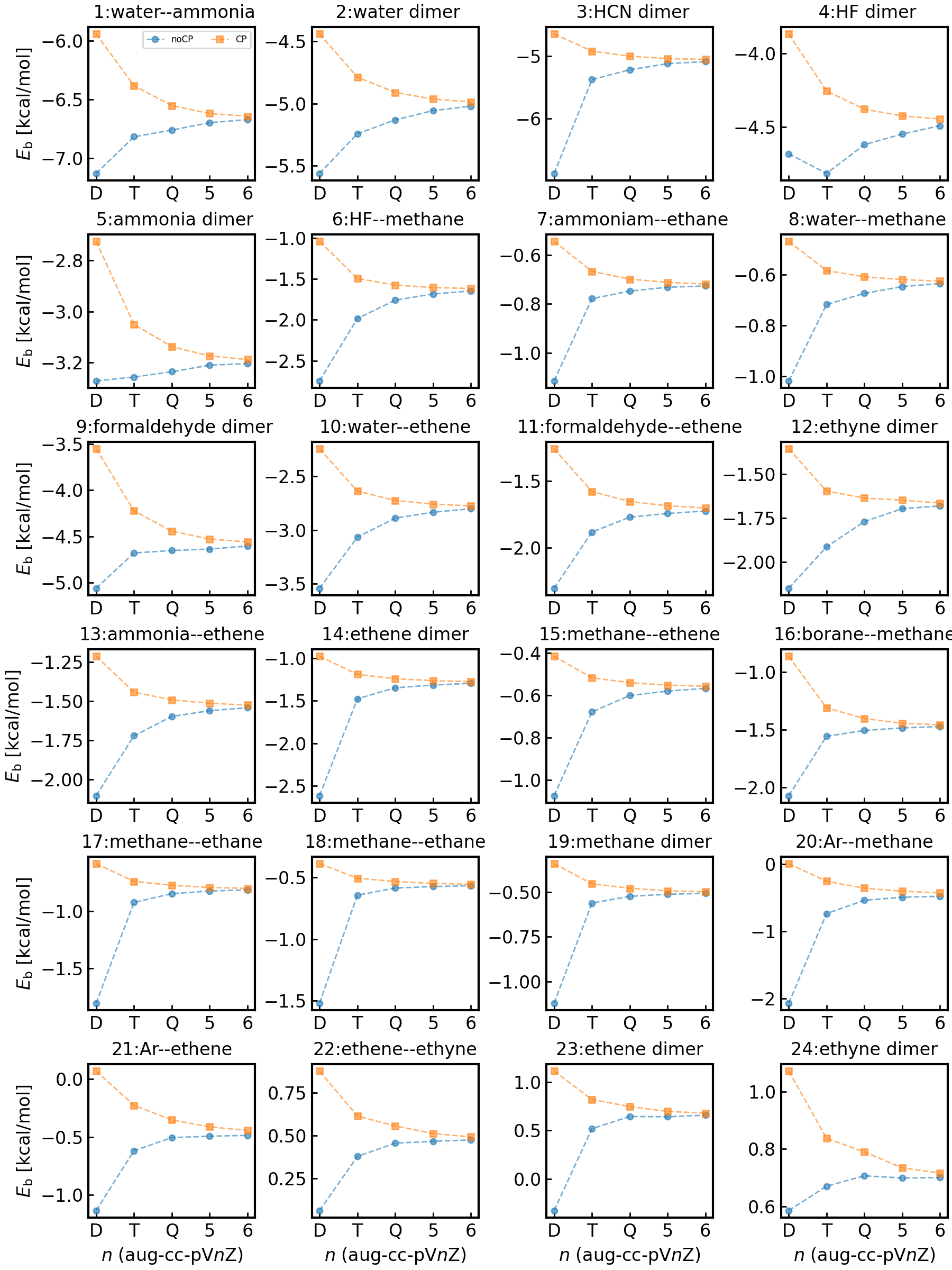}
\caption{The binding energies of A24 data set computed by MP2 with aug-cc-pV$n$Z basis sets ($n$=D,T,Q,5,6). The CP and noCP represent `with the counterpoise correction' and `without the counterpoise correction', respectively.}
\label{fig:SI-binding-mp2-augccpvnz}
\end{figure*}

\begin{figure*}[htbp]
\centering
\includegraphics[width=0.9\textwidth]{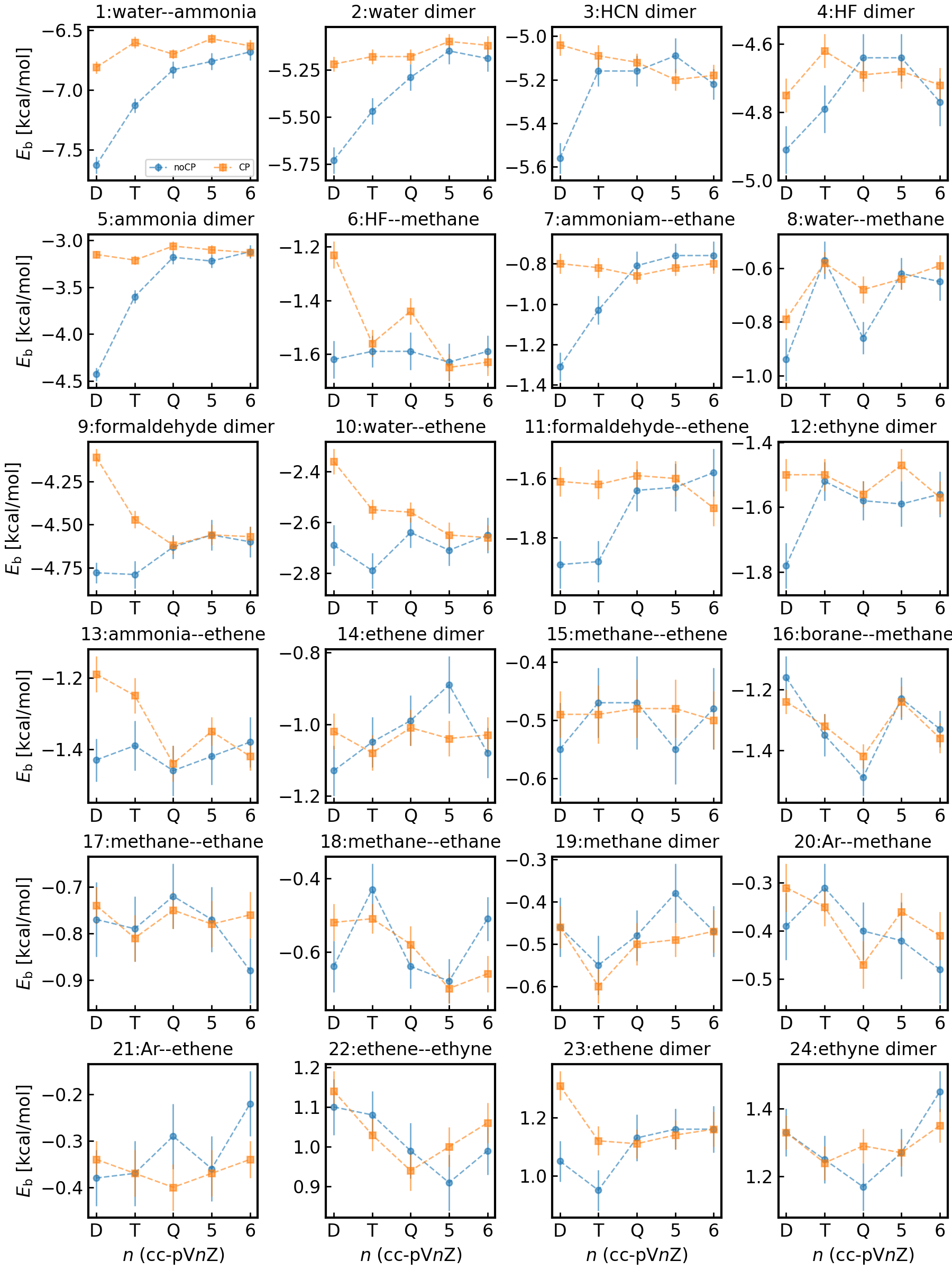}
\caption{The binding energies of A24 data set computed by LRDMC with cc-pV$n$Z basis sets ($n$=D,T,Q,5,6). The CP and noCP represent `with the counterpoise correction' and `without the counterpoise correction', respectively.}
\label{fig:SI-binding-dmc-ccpvnz}
\end{figure*}

\begin{figure*}[htbp]
\centering
\includegraphics[width=0.9\textwidth]{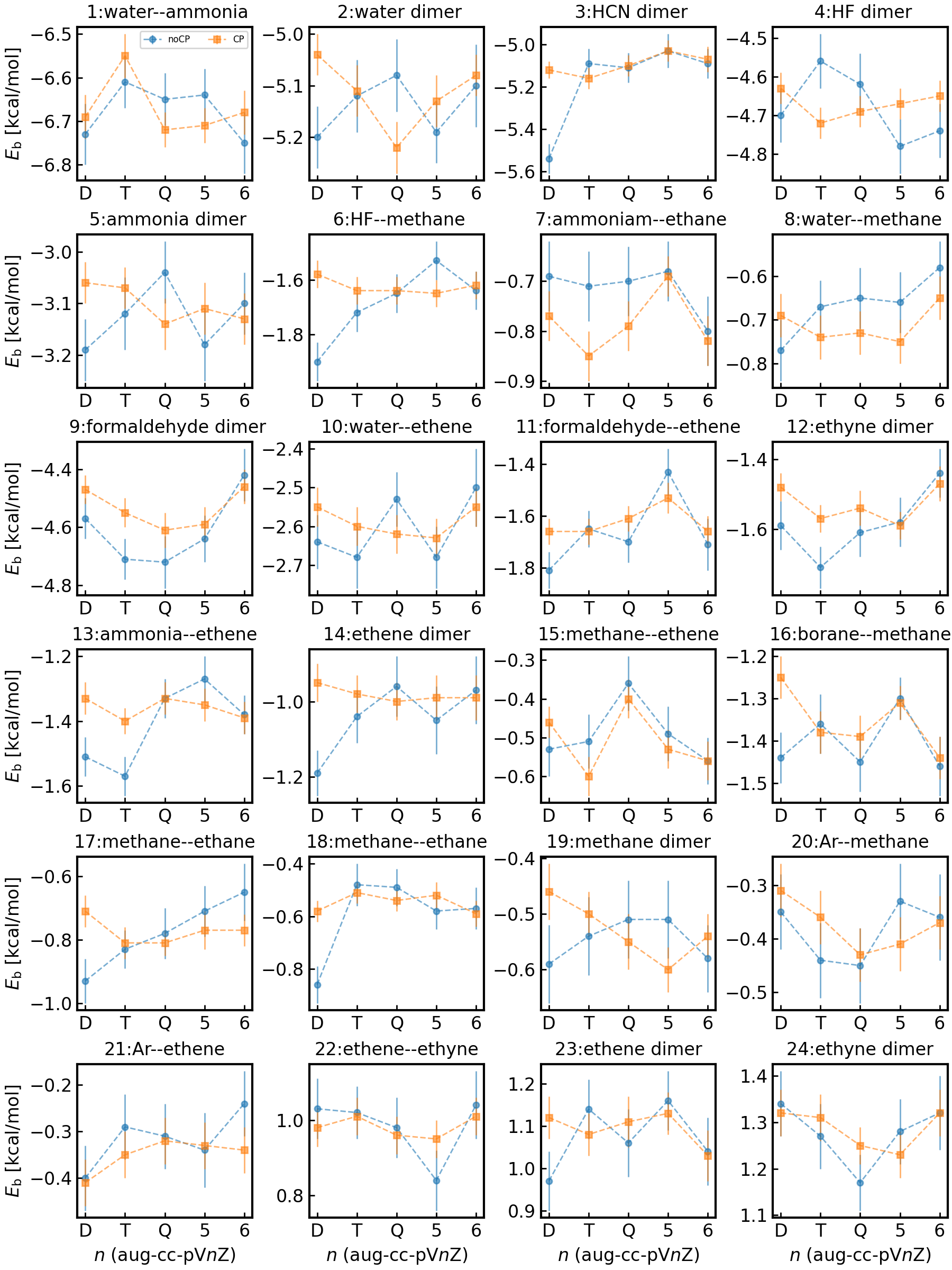}
\caption{The binding energies of A24 data set computed by LRDMC with aug-cc-pV$n$Z basis sets ($n$=D,T,Q,5,6). The CP and noCP represent `with the counterpoise correction' and `without the counterpoise correction', respectively.}
\label{fig:SI-binding-dmc-augccpvnz}
\end{figure*}

\begin{figure*}[htbp]
\centering
\includegraphics[width=1.0\textwidth]{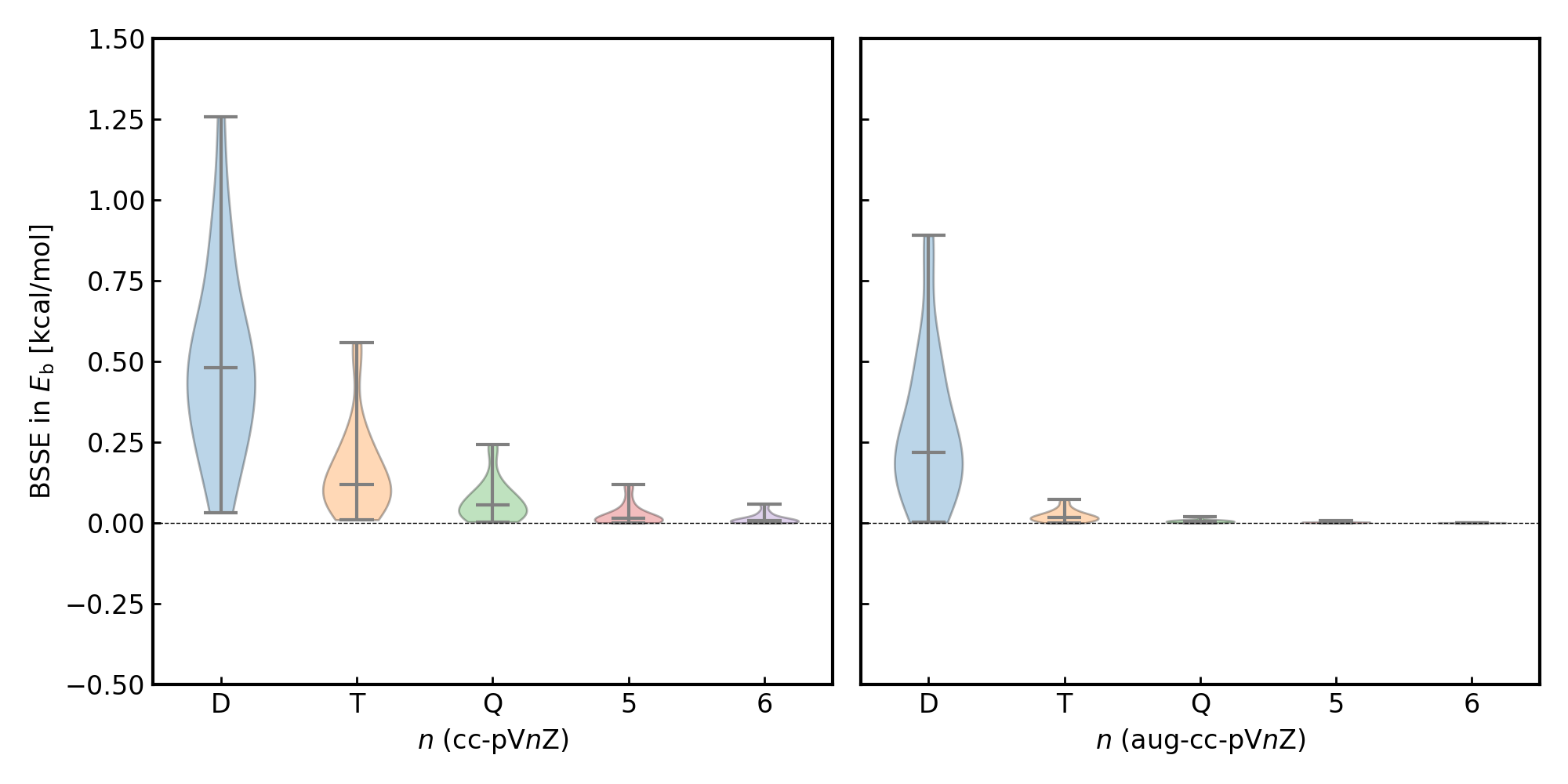}
\caption{The Violin plots of BSSEs in the binding energies of A24 data set computed by HF with (left) cc-pV$n$Z basis sets ($n$=D,T,Q,5,6) and (right) with aug-cc-pV$n$Z basis sets ($n$=D,T,Q,5,6).}
\label{fig:SI-bsse-binding-hf}
\end{figure*}

\begin{figure*}[htbp]
\centering
\includegraphics[width=1.0\textwidth]{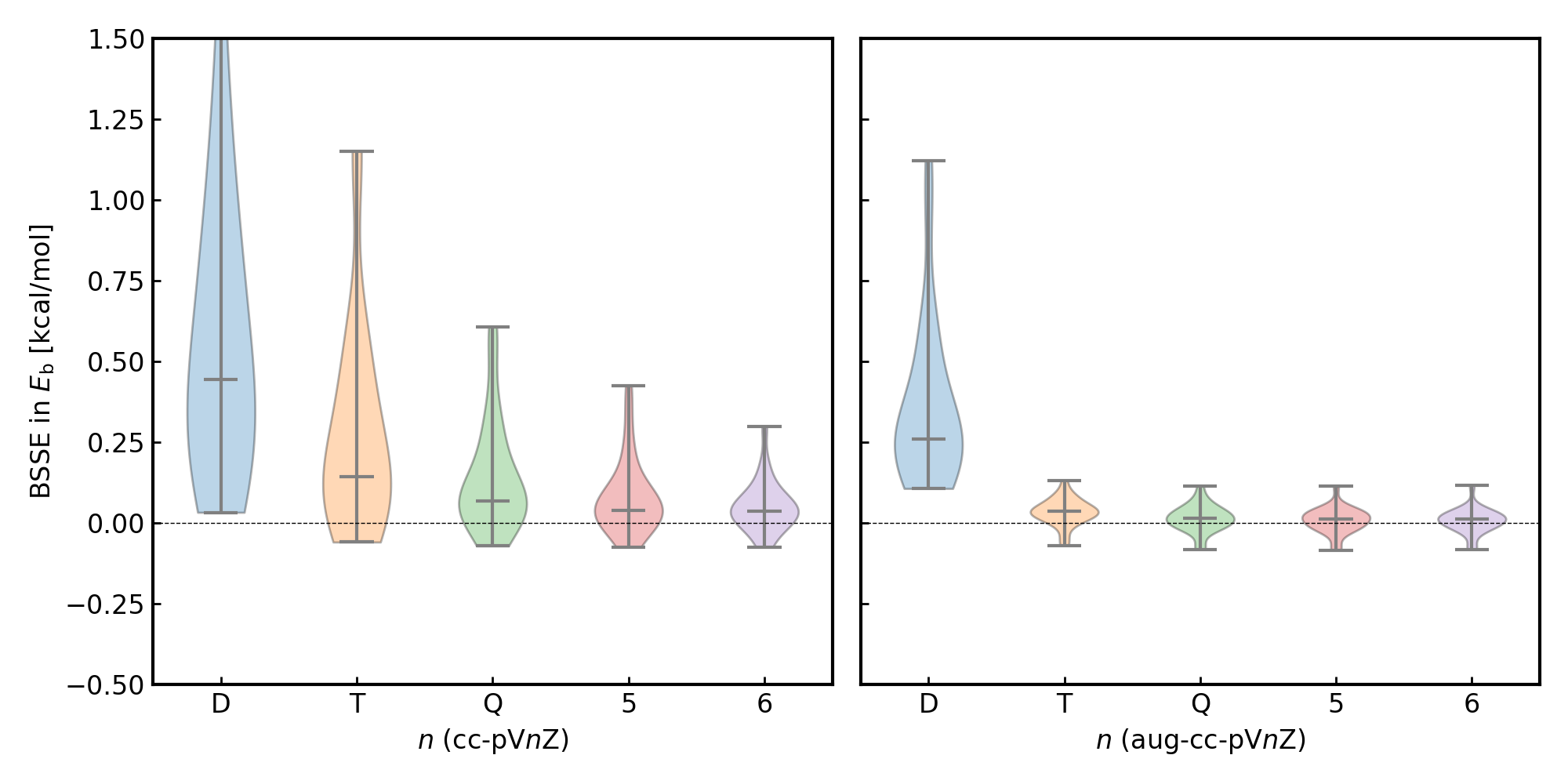}
\caption{The Violin plots of BSSEs in the binding energies of A24 data set computed by LDA with (left) cc-pV$n$Z basis sets ($n$=D,T,Q,5,6) and (right) with aug-cc-pV$n$Z basis sets ($n$=D,T,Q,5,6).}
\label{fig:SI-bsse-binding-lda}
\end{figure*}

\begin{figure*}[htbp]
\centering
\includegraphics[width=1.0\textwidth]{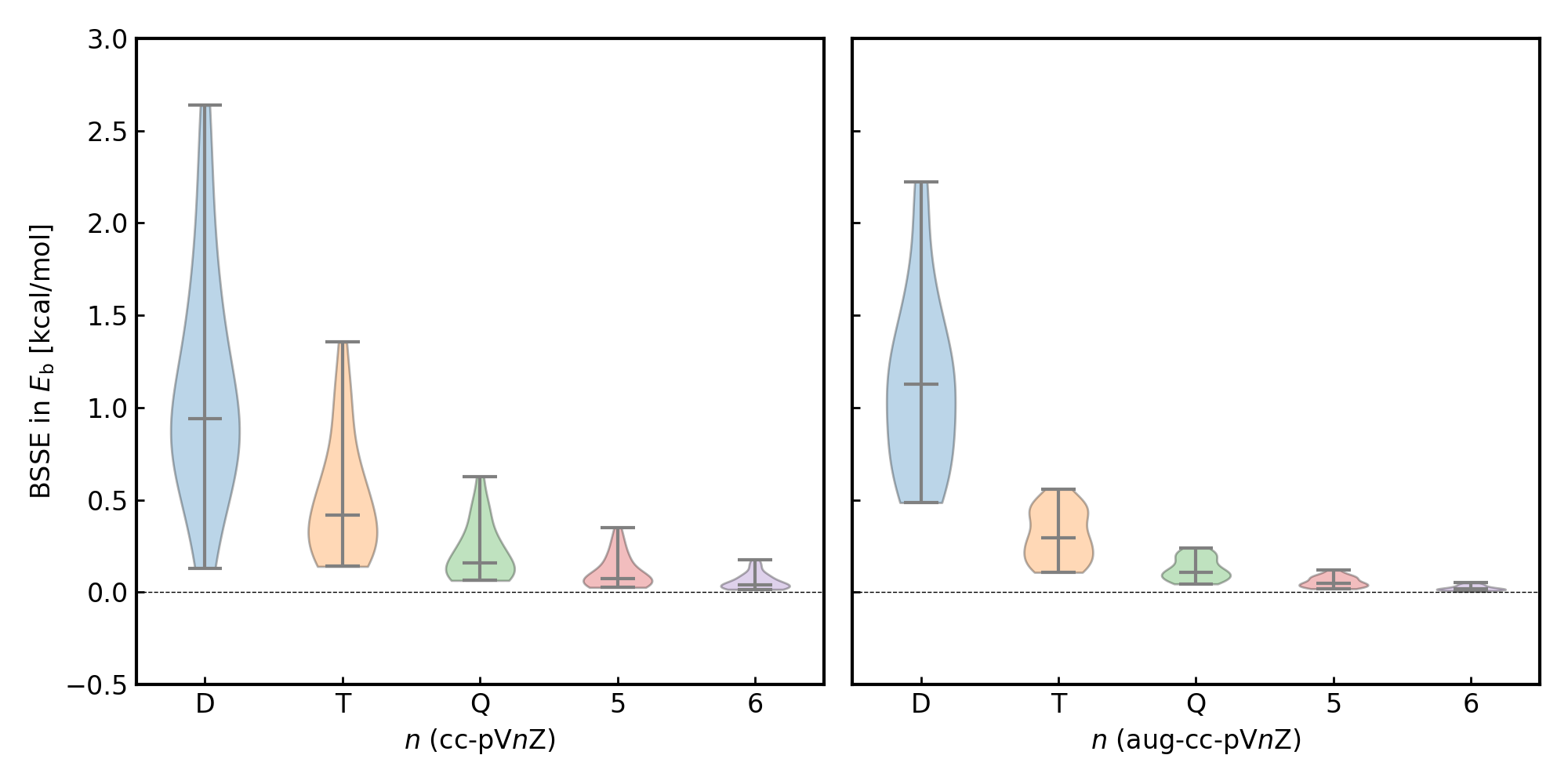}
\caption{The Violin plots of BSSEs in the binding energies of A24 data set computed by MP2 with (left) cc-pV$n$Z basis sets ($n$=D,T,Q,5,6) and (right) with aug-cc-pV$n$Z basis sets ($n$=D,T,Q,5,6).}
\label{fig:SI-bsse-binding-mp2}
\end{figure*}

\begin{figure*}[htbp]
\centering
\includegraphics[width=1.0\textwidth]{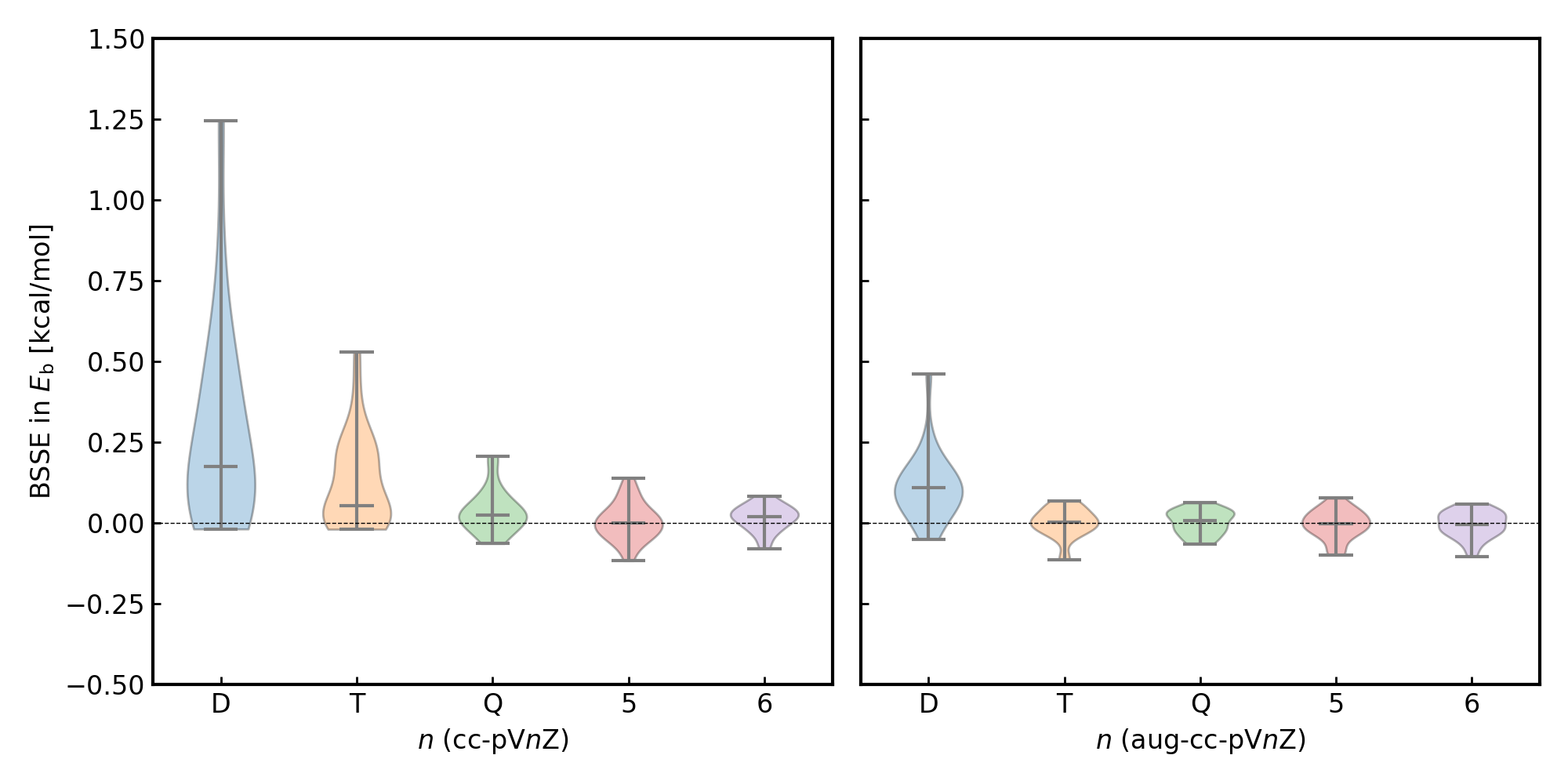}
\caption{The Violin plots of BSSEs in the binding energies of A24 data set computed by LRDMC with (left) cc-pV$n$Z basis sets ($n$=D,T,Q,5,6) and (right) with aug-cc-pV$n$Z basis sets ($n$=D,T,Q,5,6). 
}
\label{fig:SI-bsse-binding-dmc}
\end{figure*}

\begin{figure*}[htbp]
\centering
\includegraphics[width=1.0\textwidth]{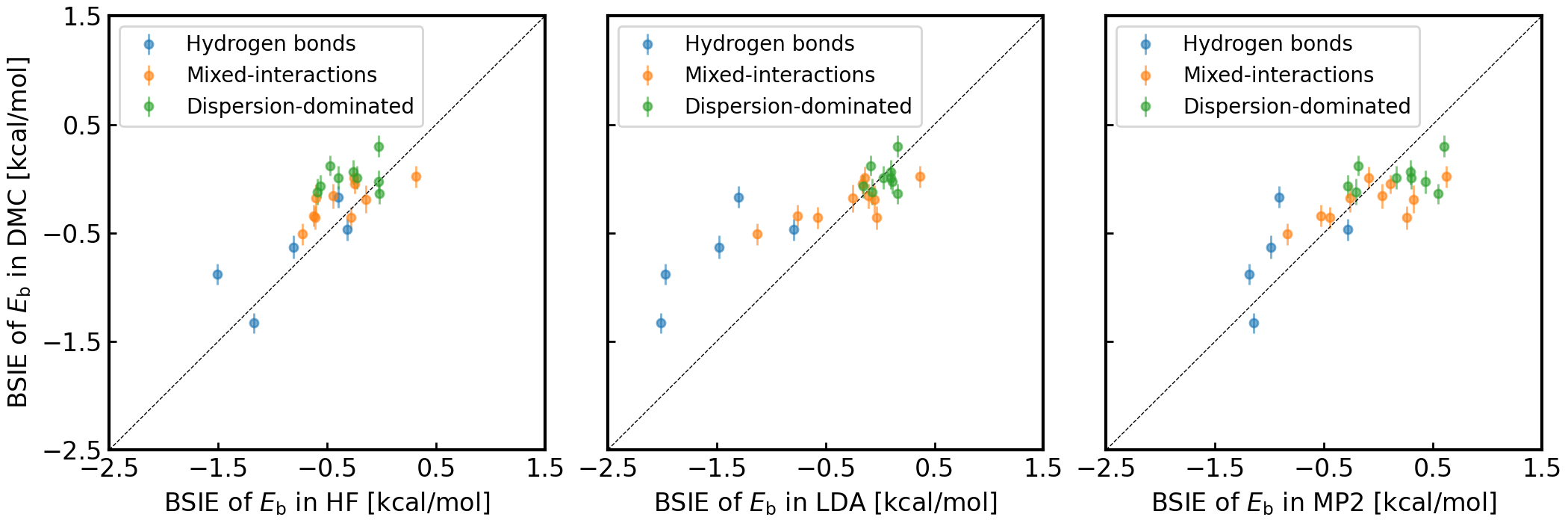}
\caption{Comparison of BSIEs in the binding energy calculations with cc-pVDZ basis set by DMC with those by HF, LDA, and MP2.}
\label{fig:SI-bsse-binding-comparison-dmc}
\end{figure*}

\clearpage
\bibliography{./references.bib}